\documentclass[10pt,floats,letterpaper]{article}
\usepackage{amsfonts,color,epsfig}
\usepackage{mathrsfs}
\usepackage{pgf}
\usepackage[latin1]{inputenc}
\usepackage{bm}
\usepackage{multirow,amssymb,amsbsy,amsmath}
\usepackage{stmaryrd}
\usepackage{epsfig}
\usepackage{epstopdf}
\usepackage{graphicx}
\usepackage{hyperref}
\usepackage{mathrsfs}

\renewcommand{\theequation}{A-\arabic{equation}}
\newcommand{\nc}{\newcommand}

\newcommand{\rnc}{\renewcommand}

\newcommand*{\bra}[1]{\mathopen{\langle}#1\mathclose{|}}
\newcommand*{\ket}[1]{\mathopen{|}#1\mathclose{\rangle}}

\makeatletter
\rnc{\theequation}{\thesection.\arabic{equation}}
\@addtoreset{equation}{section}
\makeatother
\nc{\fig}[5]{
\begin{figure}[!htbp]
    \begin{center}
    \leavevmode
    \centerline{
        \includegraphics[width=#1, height=#2]{#3}
        }
    \caption[]{#4}
    \label{#5}
    \end{center}
\end{figure}}
\nc{\figs}[8]{
\begin{figure}[!htbp]
    \begin{center}
    \leavevmode
    \centerline{
        \includegraphics[width=#1, height=#2]{#3}
        \includegraphics[width=#4, height=#5]{#6}
        }
    \caption[]{#7}
    \label{#8}
    \end{center}
\end{figure}}
\begin{document}
%\begin{flushright}
%{\tt gr-qc/0609094}
%\end{flushright}
\vspace{2mm}
\begin{center}
{{{\Large \textbf{Building all Time Evolutions with Rotationally Invariant Hamiltonians}}}}\\[16mm]
{I. Marvian\footnote{%
imarvian@sciborg.uwaterloo.ca} and R.B. Mann\footnote{%
rbmann@sciborg.uwaterloo.ca}}\\[10mm]
{\emph{Department of Physics, University of Waterloo,}\\[0pt]
\emph{Waterloo, Ontario, N2L 3G1, Canada}}\\[0pt]
{\emph{Institute for Quantum Computing, University
of Waterloo}, \\[0pt]
  \emph{Waterloo, Ontario, N2L 3G1, Canada}
}\\[0pt]
\end{center}

\vspace{8mm}
\begin{abstract}
All elementary Hamiltonians in nature are expected to be invariant
under rotation. Despite this restriction, we usually assume that any
arbitrary measurement or unitary time evolution can be implemented
on a physical system, an assumption whose validity is not obvious.
We introduce two different schemes by which any arbitrary
time evolution and measurement can be implemented with desired
accuracy by using rotationally invariant Hamiltonians that act on
the given system and two ancillary systems serving as reference
frames.  These frames specify the $z$ and $x$ directions and are
independent of the desired time evolution. We also investigate
the effects  of quantum fluctuations that inevitably arise due to usage
of a finite system as a reference frame and estimate how fast these
fluctuations tend to zero when the size of the reference frame tends to
infinity. Moreover we prove that for a
general symmetry any symmetric quantum operations can be implemented
just by using symmetric interactions and ancillas in the symmetric
states.
\end{abstract}
\vspace{10mm}

{\footnotesize ~~~~PACS numbers: }

\vspace{3cm}

\hspace{11.5cm}{Typeset Using \LaTeX}
\newpage

\section{Introduction}

In describing any physical system a reference frame is
indispensable.  The state of a system is not an abstract concept
independent of a reference frame; the only meaningful and measurable
observables are correlations between the main physical system and a
reference frame. A reference frame itself  is always represented
with a physical system.  In practice we do not consider reference
frames as quantum systems, and we treat them classically. However
there are  theoretical considerations that force us to treat
reference frames quantum mechanically. This can be regarded as the
counterpart of the famous idea that information is physical:
since information is stored in
the state of  some physical object that is subject to the laws of
physics, at the most fundamental level the abilities and limitations
of these physical objects for information processing depends on
these physical laws. In the same way, a reference frame is always
carried by  a physical system. Indeed, it has been argued that the
various conceptual conundrums in quantum mechanics -- such as controversies  about
existence of superposition between number states in quantum optics,
superconductivity and Bose-Einstein condensation, or the problem of
quantification of entanglement in systems of bosons and fermions
\cite{Spekens} -- are rooted in
ignoring the role of reference frames.

Amongst the variety of reasons for considering reference frames as
quantum mechanical objects, we are  interested in the following.
First, all elementary interactions in nature are expected to have
specific symmetries. For example, they do not have a preferred
direction and so are rotationally invariant. Given this situation, a
question arises as to whether or not it is possible to implement an
arbitrary Hamiltonian and measurements that might not have that
symmetry.  The problem of restrictions  on quantum operations imposed by
assuming a given symmetry was first emphasized by Wigner
\cite{Wigner}. Assuming all Hamiltonians commute with some  symmetry
operator, he showed  that  it is impossible to measure an observable
that does not commute with that symmetry operator. Later Araki and
Yanase \cite{Araki} proposed a scheme to measure an operator that
does not commute with that symmetry operator to any desired
accuracy. To this end they used a quantum system acting as a
reference frame beside the main system. The accuracy of that
measurement increases with the size of the Hilbert space of that
reference frame (though not necessarily linearly).

The second reason  for treating reference frames as
quantum mechanical objects, which is related to the first one,
arises from the aspiration to have a relational quantum mechanics
\cite{Rovelli91,Rovelli96}. By relational we mean a theory in which
we do not have an external reference frame that is used to define
kinematical observables.  The key motivator here is  general
relativity, a background independent theory whose quantum
generalization is also expected to inherit this property.  In the
semi-classical limit, where reference frames are assumed to have a
large Hilbert space, relational quantum mechanics reduces to
standard  quantum mechanics. However at the semi-classical limit we
cannot neglect gravitational effects due to these reference frames
\cite{Rovelli91,Rovelli96,Poulin_Toy,Poulin_REF}.  So the problem of
finding ``relational quantum mechanics" is necessarily part of the
project of constructing a quantum theory of gravity.

There are  other situations that force us to treat reference frames
as quantum mechanical objects. Specifically, in quantum computation
we need to perform measurements on a small quantum register. In this
case  the quantum register must be strongly coupled to the
apparatus. This requirement forces usage of small and cold apparatus
\cite{degrade}. For example, in some proposed experiments for
measuring spin, a small magnet is used to  measure a single spin.
Note that this magnet can be considered as the reference frame that
defines a specific direction. Due to its small size, the magnet
should be treated as a quantum mechanical object \cite{single,
single2} .

Considering a bounded quantum mechanical object as a reference frame has some side effects.
 First of all, using a bounded quantum reference frame we cannot implement the measurement or time evolution perfectly.
 There are inevitably quantum mechanical fluctuations
 \cite{Spekens,Poulin_Toy,JC}.  Second, performing a measurement or a time evolution has always some back-reaction on
 the reference frame and
 degrades it \cite{Poulin_REF,degrade,JC}.

We consider here the problem of implementing any arbitrary unitary
transformation on a physical system. We show that this can be
accomplished  using two ancillary systems that act as two reference
frames, referred to as X-RF and Z-RF, respectively specifying the x
and z directions. Note that the state of the reference frame is
independent of the desired unitary transformation. As the dimension
of the Hilbert space of both reference frames tends to infinity, the
implemented operation becomes the same as the desired one.  We
first construct a primary scheme in which, by using an isotropic
unitary transformation   and a reference frame Z-RF, we can
implement all unitary operations on the system that commute with
$L_Z$.   We then use this to construct two different schemes, \emph {scheme I and scheme II} , each
of which ensures that our desired unitary transformation is
implemented to a given accuracy.   In \emph{scheme I} we employ a
unitary transformation commuting with $L_Z$ that acts on both the
physical system and an auxiliary frame X-RF whose angular momentum
is large.  Changes in the angular momentum of the system that occur
due to the implementation of this unitary transformation  can be
compensated by changes in the angular momentum of X-RF. In the
\emph{scheme II} we construct a sequence of unitary operators that
alternately commute with $L_Z$ and  $L_X$, and respectively make use
of a sequence of Z-RFs and X-RFs, and show how any desired unitary
transformation can be implemented via this sequence.

Any measurement consists of a process in which a measurement
apparatus initially uncorrelated with a system of interest changes
so that some property of the apparatus becomes correlated with some
physical property of the system. As with all other quantum
phenomena, this can be described by a unitary time evolution acting
on the system and measurement apparatus. Therefore using this scheme
we can implement any arbitrary measurement just using isotropic
interactions; in this sense we can regard this work as a kind of
generalization of  ref. \cite{Araki}. Our work also
generalizes the simple  model  that Poulin \cite{Poulin_Toy} used to
show how it is possible to construct relational quantum mechanics
from ordinary quantum mechanics using some specific rules. In that
 toy model  a reference frame specifying the $z$
direction was constructed and used to measure the spin of a spin-1/2 particle.

Alternatively, this problem is closely related to the issue of
restrictions imposed by superselection rules. Originally these rules
were regarded as some axiomatic limitations added to the theory that
restrict the set of physically realizable operations \cite{Wick}.
For example the superselection rule for electric charge forbids
the production of superposed states with different charges. On the
other hand, as emphasized more recently  \cite{Spekens}, the lack of
a reference frame breaking some symmetry imposes restrictions
on implementable operations, which can can be regarded  as
superselection rules. For example, assuming all interactions are
rotationally invariant, without any reference frame that breaks this
symmetry, we cannot prepare states that are not rotationally
invariant; we also cannot perform  time evolutions and measurements
that are not rotationally invariant. In this way the latter
superselection rules might seem more fundamental. However as first
argued by Aharanov and Susskind  \cite{Aharanov,AharanovBook},
emergence of those fundamental superselection rules in
non-relativistic quantum theory results from making the (unphysical)
assumption that there exist absolute operations without any
dependence on some quantum reference frame. In ref. \cite{Kitaev}
Kitaev et al.  showed that all superselection rules described by
compact groups result from the lack of an appropriate reference frame.
In their terminology, an \emph{invariant world}, which is subject to
a superselection rule described by a compact Lie group by virtue of
an unbounded quantum reference frame (which itself is subject to
those superselection rules) can  simulate the physics of an
\emph{unrestricted world} that is not subject to those
superselection rules. To show this, they prove that for all states,
time evolutions, and measurements on a particular system in the
unrestricted world there exist states, time evolutions and
measurements associated with the total system, including that
particular system and quantum reference frame,  that are invariant
under the group describing those superselection rules. Moreover they
produce the same observable effects. Then using this fact they show
that superselection rules do not enhance the information-theoretic
security of quantum cryptographic protocols.

However, our work is distinct from ref. \cite{Kitaev} in that we
wish to perform an arbitrary time evolution or measurement on some
given unknown arbitrary state of a physical system; we do not want
to simulate that physics, but we want to exactly perform
unrestricted operations  using a quantum reference frame which
breaks that symmetry. We therefore  assume that the initial states
of the system and quantum reference frame are uncorrelated. However
in the scheme proposed by Kitaev et al \cite{Kitaev}, to a state of
the system in the unrestricted world they associate an initial state
in the invariant world that is not necessarily the tensor product of
that initial state of the system and a state of the reference frame.
Indeed, the only case in which this initial state is a tensor
product is  just the trivial case where the initial state of the
system is invariant under that group. Hence given some unknown
arbitrary state of the system, using their scheme we cannot perform
all quantum operations.

The preceding discussion  was concerned with cases in which
implementation of  non-symmetric time evolution requires non-symmetric
resources, i.e.  a reference frame that  breaks the symmetry. However
it is usually implicitly assumed that to implement a
 symmetric quantum operation one needs no non-symmetric resources.
In the other words, the assumption is that any symmetric quantum
operation can be realized by a symmetric unitary evolution acting on
the system and an ancillary system in a symmetric state. Before
working more on the non-symmetric cases we check the validity of
this assumption in the next section. For a general compact symmetry
group, and any given symmetric quantum operation we find an explicit
form of  a symmetric unitary time evolution acting on the system and
a symmetric ancilla that realize the given quantum
operation.\cite{Keyl}

In this paper we denote by $R$-inv a
rotationally invariant quantity, and $Z$-inv,$X$-inv for unitary
operators that are invariant under rotation around the $Z$ and $X$
axes  respectively.

The outline of our paper is as follows. In section \ref{nonsym} we
show that only symmetric resources are required to carry out any
given symmetric quantum operation. Then in section \ref{distance-sction} we
introduce a distance measure for quantum operations and investigate some of its useful properties.
In section \ref{classuni} we
discuss how to classify unitary operators with rotational symmetry,
both in general and about one axis.  We discuss in section
\ref{Z-inv-sec} how to implement $Z$-inv unitary operators using
$R$-inv (ie fully rotationally invariant) operators. In section
\ref{All} we show how to use $Z$-inv unitaries to implement all
unitary operations, and then in the next two sections we
respectively describe each of schemes I and II.  We close in section
\ref{disc} by comparing the two schemes from the point of view of the resources
they require  and with some suggestions
for further research.

\section{When are non-symmetric resources necessary?}\label{nonsym}

Suppose a unitary time evolution with a specific symmetry acts on
the main  system and an ancillary system. We assume the initial
state of ancillary system is also invariant under the symmetry. It
is easy to see that the total effect of this time evolution on the
system is described by a quantum operation that is invariant under
the symmetry. Clearly if we make use only of symmetric resources,
i.e. symmetric time evolutions and symmetric initial states, we
obtain a symmetric quantum operation.

Consider next  the inverse problem:   is it possible to  implement
any given symmetric quantum operation using only symmetric
resources, i.e. with a symmetric unitary time evolution and with an
ancillary system which initially is in a symmetric state.  As we
might guess intuitively the answer is yes as we shall now
demonstrate.

Consider a quantum operation $\varepsilon$ that is invariant under
some group described by $G$.  We call it a $G$-inv quantum
operation. Our aim is to show that for any group $G$ and for any
given $G$-inv quantum operation there exists a $G$-inv unitary time
evolution $\mathscr{S}$ acting on the system and ancilla such that
\begin{equation} \label{S-inv}
\varepsilon(\rho)=tr_{anc}(\mathscr{S} (\rho \otimes \ket{0}\bra{0})
\mathscr{S}^\dag)
\end{equation}
where $\ket{0}$, the initial state of the ancilla, is also $G$-inv.
We prove this fact and give an explicit form of such a unitary time
evolution \cite{Keyl}.

We begin by noting that it has been shown \cite{Resource} that a
$G$-inv quantum operation always admits a Kraus decomposition with
Kraus operators $K_{jm\alpha}$, where $j$ denotes an irrep, $m$ a
basis for the irrep, and $\alpha$ a multiplicity index, satisfying
\begin{equation} \label{res}
T(g) K_{jm\alpha} T^\dag(g)=\sum_{m'} u^{(j)}_{m'm}(g)
K_{jm'\alpha}\ \   \forall g\in G
\end{equation}
where $T(g)$ is the unitary operator in the Hilbert space of the system  representing  the effect of $g\in G$ on the state of system and $u^{(j)}$ is an irreducible unitary representation of $G$. We
define $\ket{j,n,\alpha}$ a basis in which ${u}^{(j)}(g)$ has the
following form.
\begin{equation}
T(g)\ket{j,n,\alpha}={u}^{(j)}(g) \ket{{j,n,\alpha}}=\sum_{n'}
{u}^{(j)}_{n'n}(g) \ket{{j,n',\alpha}}
\end{equation}
On the other hand, if $u^{(j)}$ is a representation of $G$, then complex
conjugate of $u^{(j)}$ in any specific basis is also a
representation of $G$ which might be equivalent to $u^{(j)}$. We
denote the representation obtained by complex conjugate of
${u}^{(j)}(g)$ in the basis $\ket{j,n,\alpha}$   by
$\overline{u}^{(j)}$. There  exists a basis
$\ket{\overline{j,n,\alpha}}$ such that
\begin{equation} \label{G}
T(g)\ket{\overline{j,n,\alpha}}=\overline{u}^{(j)}(g)
\ket{\overline{j,n,\alpha}}=\sum_{n'} \overline{u}^{(j)}_{n'n}(g)
\ket{\overline{j,n',\alpha}}
\end{equation}

 To purify $\varepsilon$ we assume we have an ancillary system  initially in the $G$-inv state $\ket{0}$
 as in eq. (\ref{S-inv}) . We want to build  a $G$-inv unitary $\mathscr{S}$ acting on the system and an ancillary system such that the effect of this unitary time evolution on the system is described by $\varepsilon$.  Assume $\ket{\psi}$ is an arbitrary state of system. We define $P_0$ to be the subspace spanned by all states of the form of $\ket{\psi}\ket{0}$. Also we define $P_1$ to be the subspace spanned by all states $S\ket{\psi}\ket{0}$ where $S$ is
\begin{equation} \label{P2}
S\ket{\psi}\ket{0}=\sum_{jm\alpha} K_{jm\alpha}\ket{\psi}
\ket{\overline{j,m,\alpha}}
\end{equation}
where $\ket{\overline{j,m,\alpha}}$ are states of ancillary system for
which Eq.(\ref{G}) holds.  So by definition $S$ is a map from $P_0$ to $P_1$  .We assume $\ket{0}$ is chosen such that it
is orthogonal to all states  $\ket{\overline{j,m,\alpha}}$ appearing
in Eq.(\ref{P2}). This is possible because the ancilla can always be
taken to have any number of singlet representations, one of which
can be taken to be $\ket{0}$. So obviously $P_0$ and $P_1$ are
orthogonal to each other.

Using the normalization condition $\sum_{j,m,\alpha}
K_{jm\alpha}^\dag K_{jm\alpha}=I $ it is straightforward to see
that this map preserves inner product.  Moreover we can show that
$S$ commutes with $G$. Using eqs. (\ref{res}) and (\ref{G})  we have
\begin{eqnarray}
T(g)\otimes T(g)\ S \ket{\psi}\ket{0} &=& \sum_{jm\alpha} T(g)K_{jm\alpha}\ket{\psi}\otimes T(g)\ket{\overline{j,m,\alpha}} \nonumber \\
&=& \sum_{jm\alpha} \sum_{m'} u^{(j)}_{m'm}(g) K_{jm'\alpha}
T(g)\ket{\psi} \otimes \sum_{n'} \overline{u}^{(j)}_{n'm}(g)
\ket{\overline{j,n',\alpha}} \label{unitarity}
\end{eqnarray}
and from the unitarity of $u^{(j)}$ we find
\begin{equation}
T(g)\otimes T(g)\ S \ket{\psi}\ket{0}= \sum_{jn\alpha}  K_{jn\alpha}
T(g)\ket{\psi} \otimes   \ket{j,n,\alpha}
\end{equation}
Therefore
\begin{equation} \label{commut}
(T(g)\otimes T(g))\ S \ket{\psi}\ket{0}= S  (T(g)\otimes I)
\ket{\psi}\ket{0} =S  (T(g)\otimes T(g))  \ket{\psi}\ket{0}
\end{equation}
where in the last step we have used this fact that $\ket{0}$ is
$G$-inv. So $S$ commutes with $G$. Note that $T(g)\otimes T(g)\
\ket{\psi}\ket{0}$ is still in $P_0$ and so $  T(g)\otimes T(g)
S\ket{\psi}\ket{0}$ is still in $P_1$.

 Now we define $S^{-1}$ to be a map from $P_1$ to $P_0$ such that
\begin{equation}
S^{-1}\left(\sum_{jm\alpha} K_{jm\alpha}\ket{\psi}
\ket{\overline{j,m,\alpha}}\right)=\ket{\psi}\ket{0}
\end{equation}
By definition all states in $P_1$ are of the form
$\sum_{jm\alpha} K_{jm\alpha}\ket{\psi} \ket{\overline{j,m,\alpha}}$
for some $\ket{\psi}$ and so $S^{-1}$ is defined for all states in
$P_1$. Since $S$ preserves inner product so does $S^{-1}$. Also from
Eq.(\ref{commut}) we can deduce that $S^{-1}$ commutes with $G$. We
define the subspace $P_2$ to be the subspace of all states in the
Hilbert space of system and ancillary system except those who live
in $P_0$ and $P_1$. Finally we can define a $G$-inv unitary
$\mathscr{S}$ in the following manner.
\begin{eqnarray}
\ket{\Omega} \in P_0 &:& \mathscr{S} \ket{\Omega} = S \ket{\Omega} \nonumber \\
\ket{\Omega} \in P_1 &:& \mathscr{S} \ket{\Omega} = S^{-1} \ket{\Omega}  \\
\ket{\Omega} \in P_2 &:& \mathscr{S} \ket{\Omega} = \ket{\Omega}\nonumber
\end{eqnarray}
So $\mathscr{S}$ exchanges  $P_0$ by $P_1$ and leave all states in
$P_2$ unchanged. Obviously $\mathscr{S}$ is unitary and commutes
with $G$. Also it is straightforward to check that it satisfies
Eq.(\ref{S-inv}) and so its total effect on the system would be
$\varepsilon$. This means that for any $G$-inv quantum operation one
can find a $G$-inv unitary time evolution such that  effect of this
unitary on the system is described by this quantum operation.

In general if a quantum operation is described by $N$ independent
Kraus operators, to implement it with a unitary time evolution we
need an ancillary system with an $N$ dimensional Hilbert space. In
the scheme proposed above for implementing a $G$-inv quantum
operation described by $N$ independent Kraus operators, we need an
$N+1$ dimensional ancillary. So we see that  restricting to
symmetric resources increase the minimum of size required ancillary
system at most by one.

With the same kind of argument we can prove that any given  $G$-inv
density operator of system can be purified  such that the total pure
state is $G$-inv. By  definition a $G$-inv density operator should
commute with all members of this group.

Any unitary representation of a group $G$ allows a decomposition
of the Hilbert space into charge sectors $\mathscr{H}_j$ where each charge
 sector carries an inequivalent  representation $T_j$ of $G$.
\begin{equation}
\mathscr{H}=\bigoplus_j \mathscr{H}_j
\end{equation}
Each of these sectors has $n(j)$ copies of the representation $j$. On each of these sectors the effect of
the representation of $g\in G$, $T(g)$, can be factorized to the irreducible representation associated with $j$,  $ T_j(g)$,  times an identity that acts on the multiplicity subsystem $I_{n(j)}$. Hence it can be written in the form $T_j(g) \otimes I_{n(j)} $.  Each sector can be therefore be decomposed into \emph{virtual subsystems} \cite{Zanardi}
\begin{equation}
\mathscr{H}_j=  \mathscr{M}_j \otimes  \mathscr{N}_j
\end{equation}
The effect of $T(g)$ on $\mathscr{M}_j$ is $T_j(g)$ and on $\mathscr{N}_j$ is trivial . So finally the representation $T(g)$ can be written in the form
\begin{equation} \label{rep}
T(g)=\bigoplus_{j}   T_j(g) \otimes I_{n(j)}
\end{equation}
We call  $\mathscr{M}_j$'s   \emph{gauge spaces} and   $\mathscr{N}_j$'s \emph{multiplicity spaces}. In the language of quantum information $\mathscr{M}_j$  and  $\mathscr{N}_j$ are called \emph{decoherence-full subsystems} and  \emph{noiseless} .

Using Schur's lemmas we  can show
that a $G$-inv density operator can always be written as
\begin{equation}
\rho_{G-inv}=\sum_j p_j \frac{I_j}{tr(I_j)}\otimes \rho^{(j)}
\end{equation}
where $j$ specifies different inequivalent irreps, $I_j$ is the
identity operator on the gauge subsystem of each irrep, $\rho_j$ is a density
operator acting on the multiplicity subsystem  of each irrep, and  $\{p_j\}$ is a probability distribution.
Suppose  $\ket{\phi^{(j)}}$,
which is a purification of $\rho^{(j)}$, is expanded as follows:
$$\ket{\phi^{(j)}}= \sum_{\alpha,\beta} c^{(j)}_{\alpha,\beta}\ket{\alpha}\ket{\beta}  \quad .$$
We can deduce that the following state is a $G$-inv purification of
$\rho_{G-inv}$
$$\ket{\Delta}=\sum_{j,n} \sqrt{p_j} \ket{j,n} \otimes  \ket{\overline{j,n}}\otimes \ket{\phi^{(j)}}=\sum_{j,n,\alpha,\beta} \sqrt{p_j}c^{(j)}_{\alpha,\beta} \ket{j,n,\alpha} \otimes  \ket{\overline{j,n,\beta}}$$
where $\ket{j,n}$  and    $\ket{\overline{j,n}}$ are respectivel vectors in the gauge subsystem of the main system and the ancillary one.  With the same kind of argument we
used in Eq.(\ref{unitarity}) we can see that this state is $G$-inv.
As a simple example, we can check that for the case of $SU(2)$ as
the symmetry group and the invariant density operator for spin half,
which is $I/2$,  the invariant purification state given by this
method would be the singlet state.

Suppose Alice's world is interacting and probably entangled with
Bob's world such that all of states and time evolution in her world
have a specific symmetry. According to the above discussion she can
never ascertain whether the whole world, including her system and
Bob's, has this same symmetry. On the other hand, if there existed
quantum operations or mixed states that were not purifiable by
symmetric resource, by observing those states or time evolutions she
could ascertain that the whole world is not subject to that
symmetry.

\section{Distance between two quantum operations}\label{distance-sction}

To estimate  the accuracy of implementing  unitary time evolutions
 we need a  measure to compare the implemented time evolution with the desired one.
 In this section we introduce a distance for comparing two quantum operations and investigate some of its main properties that we shall utilize in this paper.

Consider two
quantum operations $\varepsilon_1$ and $\varepsilon_2$ that map
density matrices to density matrices. By definition  these are
trace-preserving completely positive maps. Consider performing all
Von-Neumann measurements on $\varepsilon_1(\rho)$ and
$\varepsilon_2(\rho)$, and let $d(\varepsilon_1,\varepsilon_2)$ be
the maximum difference in probability between the same specified
outcomes. This comparison can be repeated for different density
matrices. We therefore define a measure
\begin{equation} \label{distance}
d(\varepsilon_1,\varepsilon_2)=\max_{P,\rho}|tr(P[\varepsilon_1(\rho)-\varepsilon_2(\rho)])|
\end{equation}
where the maximization is taken over all density operators $\rho$ and
different outcomes of all measurements that can be described by
projectors $P$. This provides a natural measure of the similarity of
two superoperators.  Clearly the distance between two quantum operations varies between zero and one.

We pause to discuss some of the properties of this distance operator.
Consider a unitary time evolution $U_N$  acting on an $N$ dimensional Hilbert space where $N$  is a large number.
 Suppose we partition this $N$ dimensional Hilbert space to a pair of subspaces with $N-2$ and 2 dimension. Assume the unitary time evolution is in the form of $I_{N-2} \oplus U_{2}$  where $I_{N-2}$ is the identity operator which acts on the $N-2$
 dimensional subspace.  Also $U_2$ acts unitarily on the 2 dimensional subspace.  So the unitary time evolution described by $U_N$ acts the same as
the  identity operator through a large  part of the $N$ dimensional Hilbert space. However, the distance of the quantum operation described by $U_N(.)U^\dag_N$ and the
 identity quantum operation can be anything between 0 and 1, dependent on the choice of element from $U_2$. Note that
 an experimentalist can easily distinguish these two quantum operations  by preparing the
 initial state of the system to
have a non-vanishing density matrix only in the 2 dimensional
subspace. So two quantum operations with large distance (as
measured by eq. (\ref{distance})) might act almost the same over a
large portion of Hilbert space. Note, however, that if the distance
of two quantum operations tends to zero, this means that they act
almost the same in all parts of Hilbert space. This property
justifies the usage of this specific distance for our purpose. If
the distance of a simulated quantum operation with the desired one
tends to zero then, no matter what is the size of Hilbert space, the
simulated quantum operation acts exactly the same as the desired one
in all parts of Hilbert space and so they are indistinguishable.

We may also express this definition by making
use of the \emph{trace distance} $D(\rho_1,\rho_2)$ between two density operators \cite{Nielsen}
$$d(\varepsilon_1,\varepsilon_2)=\max_{\rho} D(\varepsilon_1(\rho),\varepsilon_2(\rho))$$
One of the important properties of trace distance is  $D(\varepsilon(\rho_1),\varepsilon(\rho_2)) \leq D(\rho_1,\rho_2)$ \cite{Nielsen}. Using this property and the triangle inequality  we can deduce
\begin{equation} \label{trace_distance}
d(\varepsilon_1(\varepsilon_2(.)),\varepsilon_1^{'}(\varepsilon_2^{'}(.)))\leq
d(\varepsilon_1(.),\varepsilon^{'}_1(.))+d(\varepsilon_2(.),\varepsilon_2^{'}(.))
\end{equation}
Suppose $\rho_0$ is the density operator and $P_0$ the projector for
which this maximum in Eq.(\ref{distance}) occurs. Let $\{\ket{i}\}$
be a basis in which $\varepsilon_1(\rho_0)-\varepsilon_2(\rho_0)$ is
diagonal. Dividing
 $\{\ket{i}\}$ into two distinct eigenspaces $\{\ket{i}^{+}\}$   and $\{\ket{i}^{-}\}$
 of positive and negative eigenvalues, we denote the projector to these subspaces by $P^+$
 and $P^-$. The trace of $\varepsilon_1(\rho_0)-\varepsilon_2(\rho_0)$ is zero, so obviously
\begin{equation}
|tr(P^-[\varepsilon_1(\rho_0)-\varepsilon_2(\rho_0)])|=|tr(P^+[\varepsilon_1(\rho_0)-\varepsilon_2(\rho_0)])|=\frac{1}{2}\sum_i
|\bra{i}(\varepsilon_1(\rho_0)-\varepsilon_2(\rho_0))\ket{i}|
\end{equation}
Consequently the projector $P_0$, for which Eq.(\ref{distance}) is
maximum, can be chosen to be either of $P^+$ or $P^-$. Moreover
\begin{equation}  \label{distance2}
d(\varepsilon_1,\varepsilon_2)=\frac{1}{2} \max_{\rho,\{\ket{i}\}}
\sum_i |\bra{i}(\varepsilon_1(\rho)-\varepsilon_2(\rho))\ket{i}|
\end{equation}
where the maximization is taken over all density operators and all
orthogonal bases.  Using the preceding equation we can readily show the convex property
\begin{equation}  \label{convex}
d(\varepsilon,\sum_i p_i \varepsilon_i)\leq \sum_i p_i d(\varepsilon, \varepsilon_i)
\end{equation}

Suppose $\varepsilon^{A,B}$ is a quantum operation acting on systems $A$ and $B$. Assume initially the  state of the
total system is $\rho^A \otimes \rho^B$  where $ \rho^B$ is some fixed state. The effect of $\varepsilon^{A,B}$ on system $A$ is then given by $\varepsilon^{A}(\rho^A)\equiv tr_B(\varepsilon^{A,B}(\rho^A \otimes \rho^B))$. Using Eq.(\ref{distance2}) it is straightforward to check that
\begin{equation} \label{distanceproperty}
d(\varepsilon^{A}_1(.),\varepsilon^{A}_2(.) )\leq
d(\varepsilon^{A,B}_1(.),\varepsilon^{A,B}_2(.) )
\end{equation}
This  inequality expresses the intuitive notion that distinguishing between two different quantum operations is more
difficult when one observers only part of a larger system.

The following Lemma (proved in the appendix) provides a useful tool for computing $d(\varepsilon_1,\varepsilon_2)$.

\textbf{Lemma I:} \emph{Suppose $\rho$ is a density operator and
$O_1, O_2$ are two arbitrary operators and $\{\ket{i}\}$ is an
arbitrary set of an orthogonal basis. Then}
$$\sum_i |\bra{i}O_1 \rho O_2 \ket{i}| \leq ||O_1||\times ||O_2||$$

Here we have used the infinity norm of an operator, namely
\begin{equation}
||\mathscr{O}||=\max_{\ket{\psi}} |\mathscr{O} \ket{\psi}|
\end{equation}
where the maximization is taken over all normalized vectors
$\ket{\psi}$. Note that $ ||\mathscr{O}_1 \mathscr{O}_2|| \leq
||\mathscr{O}_1||\ ||\mathscr{O}_2||$.
Using Eq.(\ref{distance2}) and  lemma I we can easily see that
\begin{equation} \label{unitary distance}
d(U(.)U^{\dag},V(.)V^{\dag})\leq ||U-V||+\frac{1}{2} ||U-V||^2
\end{equation}
So for small $||U-V||$ we have $d(U(.)U^{\dag},V(.)V^{\dag})\leq ||U-V||$.

\section{Classification of unitaries with rotational
symmetries}\label{classuni}

As we saw in the  section 2 , the general form of a unitary
representation of a  group is given by Eq.(\ref{rep}). For the
rotation group angular momentum is the related index that specifies
different representations. So the effect of an  arbitrary rotation
$R$ on a physical system can be represented by a unitary matrix
$T(R)$ that (reducibly) decomposes as
\begin{equation}
T(R)=\bigoplus_{l_{min}}^{l_{max}} T_l(R) \otimes  I_{n(l)}
\end{equation}
where $T_l(R)$ is the irreducible representation (irrep) of a
rotation with angular momentum $l$. Here $n(l)$ is the multiplicity
of this angular momentum and $I_{n(l)}$ is the $n$ dimensional
identity.  We say that a unitary transformation is rotationally
invariant  iff
\begin{equation}
[U,T(R)]=0
\end{equation}
for all rotations $R$. We denote rotationally invariant unitary
operators by  $U_{\mathscr{R}{\rm -inv}}$. Noting that $T_l(R)$ is
an irrep of the rotation group, we deduce using Schur's lemmas that
all possible rotationally invariant unitaries  have the form
\begin{equation} \label{unitary}
U_{\mathscr{R}{\rm -inv}}=\bigoplus_{l_{min}}^{l_{max}} I_{l} \otimes
U^{(l)}
\end{equation}
where $U^{(l)}$ is an arbitrary unitary operator acting on the
$n(l)$-dimensional  Hilbert space of multiplicity subsystems and $I_{l}$ is the identity
operator on the gauge subsystems.

Now consider a state $\ket{j,m,\lambda}$ with angular momentum $j$,
magnetic quantum number $m$, and $\lambda$ some other possible
quantum number. The effect of $U_{\mathscr{R}{\rm -inv}}$ on this
state is
\begin{equation}
U_{\mathscr{R}{\rm -inv}}\ket{j,m,\lambda}=\sum_{\lambda'}
U_{\lambda' \lambda}^{(j)} \ket{j,m,\lambda'}
\end{equation}
where $U_{\lambda' \lambda}^{(j)}$ acts unitarily on the  subspace
of $\lambda$-multiplets with the same $j$ and $m$.

Consider next all unitary operators $V$ that  have one axis of
symmetry.  Without loss of generality we can take this to be the
$z$-axis, ie. we consider all unitaries that commute with $L_z$.
They can be decomposed via
\begin{equation} \label{Z-inv}
V_{\mathscr{Z}{\rm -inv}}=\bigoplus_M \
V^{(M)}=\sum_{M,\lambda,\lambda'} V^{(M)}_{\lambda,\lambda'}
\ket{M,\lambda}\bra {M,\lambda'}
\end{equation}
where $\{\ket{M,\lambda}\}$ for different $\lambda$ is an orthogonal
basis for the subspace where $L_Z=M$. The operator $V^{(M)}$ acts
unitarily on this subspace.

\section{Implementing $Z$-inv unitaries using $R$ -inv unitaries } \label{Z-inv-sec}

In this section, for any given unitary time evolution with
one axis of symmetry,  a $Z$-inv unitary, we construct a rotationally invariant unitary
time evolution such that, when this $R$-inv unitary acts on the
combined physical system with Z-RF, the total effect on the physical
system is equivalent to that of the given $Z$-inv unitary. As noted
in the introduction we refer to this procedure as implementation of
a $Z$-inv unitary.

The main idea  behind this construction is based on the simple
property that when two vectors of unequal norm are added together,
the length of the resultant vector is almost independent of the
components of the vector of smaller norm that are orthogonal to the
larger vector, provided the former has sufficiently small norm (see fig. 1).
Lemma II demonstrates the quantum version of this simple property.

%\fig{10.5cm}{5cm}{add.eps}
%{\small Adding a large vector to a small one: The length of the sum,  $A_1$ and $A_2$ of the  large vector $L$ and the small vectors $S_1$ and $S_2$ is
% almost independent of the components of $S_1$ and $S_2$ which  is perpendicular
%to the large one. Lemma II demonstrates the quantum version of this simple property.}
%{fig:add}

\begin{figure}%[h!]
\begin{center}
\includegraphics[width=.75 \columnwidth]{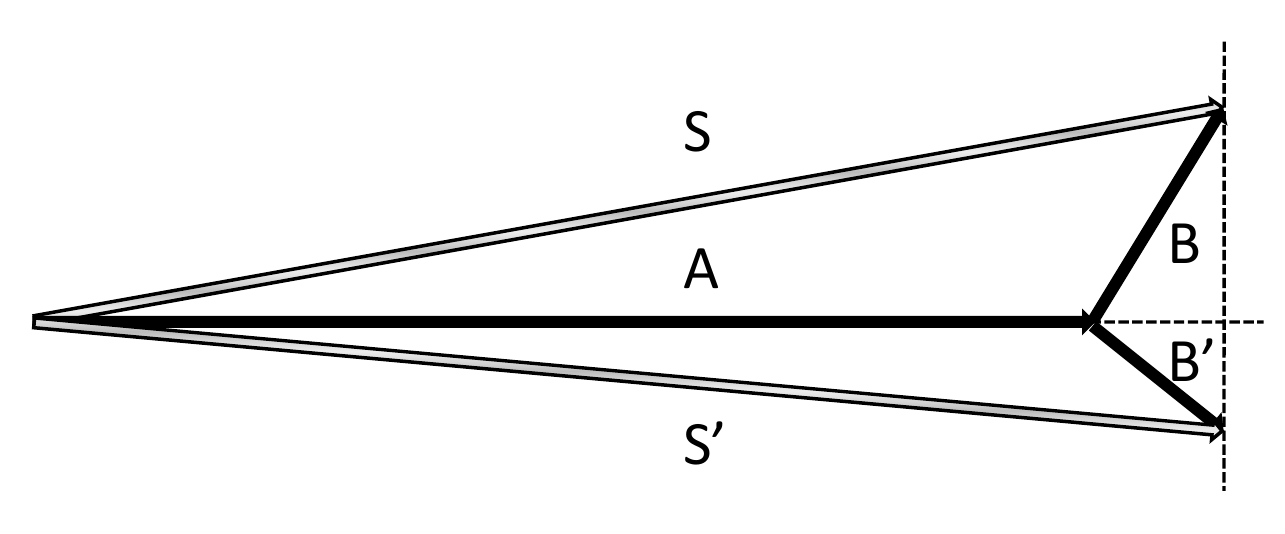}
\caption{Adding a large vector to a small one: The length of the sums,  $S$ and $S'$ of the large vector $A$ and the small vectors $B$ and $B'$ is almost independent of the components of $B$ and $B'$ which  is perpendicular
to $A$. Lemma II demonstrates the quantum version of this simple property. }
\end{center}
\label{add}
\end{figure}

%\begin{figure}
%\begin{center}
%\leavevmode
%\epsfxsize=4in
%\includegraphics[width=3.5in, height=1.3in]{add.eps}
%\epsfbox{add.eps}
%\end{center}
%\caption{ \textbf{Adding a large vector to a small one: The length of the sums,  $S$ and $S'$ of the large vector $A$ and the small vectors $B$ and $B'$ is almost independent of the components of $B$ %and $B'$ which  is perpendicular
%to $A$. Lemma II demonstrates the quantum version of this simple property. }}

\textbf{Lemma II:} \emph{Consider the following state
\begin{equation}
\ket{\psi}=\ket{l_1,m_1}\ket{l_2,m_2=l_2-k}
\end{equation}
where $l_1,l_2$ are angular momenta and $m_1,m_2$ are eigenvalues
of $L_Z$. At the limit of  $l_2\gg l_1^2, k^2$ this state is almost
the same as the state with total angular momentum equal to $m_1+l_2$
and total $L_z$ equal to $m_1+l_2-k$ i.e.
\begin{equation}
\ket{\phi}=\ket{(j=m_1+l_2,m=m_1+l_2-k):(l_1,l_2)}
\end{equation}
or more precisely
\begin{equation}
|\bra{\phi}\psi\rangle|^2 \geq 1-\frac{(l_1^2+l_1-m_1^2)(2k+1)-m_1
}{2l_2}
\end{equation}
In terms of Clebsch-Gordon coefficients this means}
$$lim_{l_1^2/l_2, k^2/l_2\rightarrow 0} \left( C_{l_2,l_2-k;l_1,m_1}^{m_1+l_2-k,m_1+l_2}\right)^2
= 1-\frac{(l_1^2+l_1-m_1^2)(2k+1)-m_1 }{2l_2} $$

This lemma is proven in the appendix. A specific case of these
relations is reported in \cite{Poulin_Toy}  where it is verified
numerically.

For $k=0$ we have
\begin{equation} \label{lemmaI}
|\bra{\phi}\psi\rangle|^2 \geq 1-\frac{l_1^2+l_1-m_1^2-m_1 }{2l_2}
\geq 1-C^2
\end{equation}
where $C^2=\frac{l_1^2+l_1+1/4}{2l_2}$ is the minimum of
$1-\frac{l_1^2+l_1-m_1^2-m_1 }{2l_2}$, which occurs for
$m_1=-1/2$.

Now consider an ancillary system called Z-RF with angular momentum
$l_{RZ}$ that is large in comparison with the maximum angular
momentum of system $l_1$. We assume initially Z-RF is in the state
$\ket{\rm{Z-RF}}\equiv \ket{l_{RZ},m_{RZ}=l_{RZ}}$ and so has
maximal angular momentum in the $z$ direction. Following the proof
in lemma II, to a very good approximation the total angular momentum
of the combined system with Z-RF depends only on $m_1$ (the $z$
component of angular momentum of the system)
\begin{equation}
\ket{l_1,m_1,\delta}\ket{{\rm Z-RF}} \approx
\ket{j=m_1+l_{RZ},m=m_1+l_{RZ}, \lambda}
\end{equation}
and so is almost independent of $l_1$.  Here $\delta$ labels
possible degeneracies of the state of the physical system and
$\lambda$ labels all distinct states of the total system with the
same $j=m_1+l_{RZ}$ and $m=m_1+l_{RZ}$. More precisely $\lambda$
indicates from which $(l_1,m_{RZ},l_{RZ},\delta)$ the state
$\ \ket{j=m_1+l_{RZ},m=m_1+l_{RZ}, \lambda}$ is formed. Since
$m_{RZ}$ and $l_{RZ}$ are fixed, different $\lambda$'s stand for
different $(l_1,\delta)$'s.

The upshot is that $j$ and $m$ of the combined system are
independent of $(l_1, \delta)$, and depend only on $m_1$.  Hence we
can implement any arbitrary unitary on the subspace of states with
the same $m_1$ by using an appropriate $U_{\mathscr{R}{\rm -inv}}$.
In other words we can  implement all unitary time evolutions on the
system that commute with $L_z$ acting on the system.   In the
language of quantum information, we  implement an arbitrary Z-inv
unitary on the system by performing an $R$-inv unitary on the
noiseless subsystem of the total system, which consists of the main
system and reference frame.

Suppose we are going to implement a unitary time evolution operator
$V$ on the system that commutes with $L_Z$. $V$ can be written in
the form of Eq.(\ref{Z-inv}).

Now consider the following decomposition of total Hilbert space of system and Z-RF induced by rotational group
\begin{equation}
\mathscr{H}_{sys} \otimes \mathscr{H}_{Z-RF}  \longrightarrow \bigoplus_j\   \mathscr{M}_j \otimes  \mathscr{N}_j
\end{equation}

We define the unitary $\mathscr{V}$
acting on the system and Z-RF to be
\begin{equation}
\mathscr{V}=\bigoplus_{j}  I_j\otimes V^{(j-l_{RZ})}
\end{equation}
where $ I_j$ acts on the gauge subsystem $ \mathscr{M}_j$, and $V^{ (j-l_{RZ}) }$ acts on the multiplicity subsystem
$\mathscr{N}_j$.
Obviously  $\mathscr{V}$ has the same form as (\ref{unitary}) and so is isotropic. Note
that here we assume this unitary governs the total system (ie. the
physical system combined with Z-RF), and so the context of the above
formula is for the total system. We assume this unitary acts on the
initial state of the total system, which is a tensor product of the
initial state of the system and $\ket{Z-RF}$.

The effect of this time evolution on the system is described by a
superoperator called $\varepsilon^{(1)}$. One possible
representation of this time superoperator can be specified by the
following set of Kraus operators
\begin{equation}
K_{n}=\bra{l_{RZ},n} \mathscr{V} \ket{Z-RF} = \bra{l_{RZ},n}
\mathscr{V} \ket{l_{RZ},l_{RZ}}
\end{equation}
where $\ket{l_{RZ},n}$ is the eigenvector of $L_Z$ with eigenvalue
$n$. For $K_{l_{RZ}}$ we  have
$$K_{l_{RZ}}=\bra{l_{RZ},l_{RZ}} \mathscr{V} \ket{l_{RZ},l_{RZ}}=\sum_{j,\lambda,\lambda'} V^{(j-l_{RZ})}_{\lambda,\lambda'} \langle l_{RZ},l_{RZ} \ket{j,j,\lambda}\bra{j,j,\lambda'}l_{RZ},l_{RZ}\rangle$$
\begin{equation}\label{K_L}
+\sum_{j,M<j,\lambda,\lambda' } V^{(j-l_{RZ})}_{\lambda,\lambda'}
\langle l_{RZ},l_{RZ}
\ket{j,M,\lambda}\bra{j,M,\lambda'}l_{RZ},l_{RZ}\rangle
\end{equation}
We can rewrite the second term as
$$V'\equiv \bra{l_{RZ},l_{RZ}}(\mathscr{I}-\mathscr{P}) \mathscr{V}(\mathscr{I}-\mathscr{P})\ket{l_{RZ},l_{RZ}}$$
where $\mathscr{I}$ is the identity operator in the Hilbert space of
the combined system and RF, and $\mathscr{P}$ is the projector on
the space of all vectors with $M=j$. So the operator
$\mathscr{I}-\mathscr{P}$ is the projector to the subspace of states
with $M<j.$ For an arbitrary  normalized vector $\ket{\Theta}$ in
the Hilbert space of the physical system  we know from
Eq.(\ref{lemmaI}) that
$|(\mathscr{I}-\mathscr{P})\ket{\Theta}\ket{l_{RZ},l_{RZ}}|\leq C$
where $C^2=\frac{l_1^2+l_1+1/4}{2l_{RZ}}$. Since $\mathscr{V}$ is
unitary we deduce that   $||V'||\leq C^2$.

To compute the first term in Eq.(\ref{K_L}) we note that
\begin{equation} \label{inner}
\bra{j,j,\lambda}(\ket{l_1,m,\delta}\otimes\ket{l_{RZ},l_{RZ}})=\delta_{j,m+l_{RZ}}
\delta_{\lambda,(l_1,\delta)} \times \xi_{m,\lambda}
\end{equation}
where $\xi_{m,\lambda}$ is some real number. From Eq.(\ref{lemmaI})
we deduce that $\xi_{m,\lambda}^2\geq 1-C^2$  and so
$\xi_{m,\lambda}^2$ is close to one. So using Eq.(\ref{inner}) we
have
\begin{equation} \label{Clebsh}
\langle l_{RZ},l_{RZ} \ket{j,j,\lambda}=\xi_{j-l_{RZ},\lambda}\
\ket{m=j-l_{RZ},\lambda}
\end{equation}
Note that $\lambda$ specifies $l_1$ and $\delta$. Now using this
equality we see that the first term in Eq.(\ref{K_L}) can be
rewritten as
\begin{equation} \label{first}
\sum_{j,\lambda,\lambda'} V^{(j-l_{Rz})}_{\lambda,\lambda'} \langle
l_{RZ},l_{RZ}
\ket{j,j,\lambda}\bra{j,j,\lambda'}l_{RZ},l_{RZ}\rangle=\sum_{m,\lambda,\lambda'}
V^{(m)}_{\lambda,\lambda'} \xi_{m,\lambda}{\xi_{m,\lambda'}}
\ket{m,\lambda}\bra{m,\lambda'}
\end{equation}

To compute this, first we define
\begin{equation}
X=\langle l_{RZ},l_{RZ}|\mathscr{I}-
\mathscr{P}|l_{RZ},l_{RZ}\rangle=I-\sum_{m,\lambda}
\xi_{m,\lambda}^2 \ket{m,\lambda}\bra{m,\lambda}
\end{equation}
To get the last equality we have used Eq.(\ref{Clebsh}). Note that
$\xi_{m,\lambda}^2\geq 1-C^2$ and so $||X||\leq C^2$ . Since
$\xi_{m,\lambda}$ is close to one we can easily see that
$\xi_{m,\lambda}\approx 1/2(1+\xi_{m,\lambda}^2)$. So we can see
Eq.(\ref{first}) equals
\begin{equation}
(I-\frac{X}{2})V(I-\frac{X}{2})\approx V- \frac{1}{2}(VX+XV)
\end{equation}
So finally $K_{l_{RZ}}=V+\overline{V}$ where $\overline{V} \equiv
V'-\frac{1}{2}(XV+VX)$. By the triangle inequality we can see that
$||\overline{V}||\leq 2C^2$.  Finally we obtain
\begin{equation}
{\varepsilon^{(1)}}(\rho)\approx V \rho V^\dag +
\varepsilon^{(1)}_{noise}(\rho)
\end{equation}
where
\begin{equation}
 \varepsilon^{(1)}_{noise}(\rho)=(V\rho {\overline{V}}^\dag+{\overline{V}} \rho V^\dag)+\sum_{n<l_{RZ}} K_{n}\rho {K_{n}}^\dag
\end{equation}
Note that $\varepsilon^{(1)}_{noise}(\rho)$ is not necessarily a
positive operator. The first two terms in
$\varepsilon^{(1)}_{noise}(\rho)$ commute with $L_Z$ and are
responsible for the noise in each block of the same $m$. The effect
of the last term is that of mixing states with different $m$.

While our target was implementing a unitary time evolution described
by $V$,  we have instead  implemented a time evolution that is
described by $\varepsilon^{(1)}$. To estimate the quality of this
implementation we compute the distance of these two time evolutions,
$d(V(.)V^\dag,\varepsilon^{(1)}(.))$:
\begin{equation}
d(V(.)V^\dag,\varepsilon^{(1)}(.))=\frac{1}{2}
\max_{\rho,\{\ket{i}\}}\sum_i |\bra{i}
\varepsilon^{(1)}_{noise}(\rho) \ket{i}| =\frac{1}{2} \max(\sum_i
|\bra{i} (V\rho {\overline{V}}^\dag+{\overline{V}} \rho V^\dag)+
\sum_{n<l_{RZ}} K_{n}\rho {K_{n}}^\dag \ket{i}|)
\end{equation}
where the maximization is over all bases $\{\ket{i}\}$ and all
density operators $\rho$. To calculate this quantity first we note
that $\sum_{n<l_{RZ}} K_{n}\rho {K_{n}}^\dag$ is a positive operator
and so
\begin{eqnarray}
\sum_i |\bra{i}\sum_{n<l_{RZ}} K_{n}\rho {K_{n}}^\dag \ket{i}| &=& tr(\sum_{n<l_{RZ}} K_{n}\rho {K_{n}}^\dag) \nonumber \\
&=&1- tr(K_{l_{RZ}} \rho {K_{l_{RZ}}}^\dag) \nonumber \\
&=&1-tr((V+\overline{V})\rho(V+\overline{V})^\dag) \nonumber \\
&=& -tr(V\rho \overline{V}^\dag+ \overline{V}\rho V^\dag
+\overline{V}\rho \overline{V}^\dag )
\end{eqnarray}
For any set of operators $\{O_i\}$ we have $|tr(O_1\ldots O_N \rho)|
\leq ||O_1|| \ldots ||O_N||$. Using this fact and noting that
$||\overline{V}||\leq 2C^2$ we can show that for small $C^2$ we have
\begin{equation}
tr(\sum_{n<l_{RZ}} K_{n}\rho {K_{n}}^\dag) = |tr(V\rho
\overline{V}^\dag+ \overline{V}\rho V^\dag +\overline{V}\rho
\overline{V}^\dag )| \leq 4 C^2
\end{equation}
On the other hand, we have
\begin{equation}
\sum_i |\bra{i} (V\rho {\overline{V}}^\dag+{\overline{V}} \rho
V^\dag) \ket{i}| \leq \sum_i |\bra{i} V\rho {\overline{V}}^\dag
\ket{i}|+ \sum_i|\bra{i}{\overline{V}} \rho V^\dag)\ket{i}|
\end{equation}
Using lemma I we have
\begin{equation}
\sum_i |\bra{i} (V\rho {\overline{V}}^\dag+{\overline{V}} \rho
V^\dag) \ket{i}| \leq 4C^2
\end{equation}
and so
\begin{equation} \label{e1}
d(V(.)V^\dag,\varepsilon^{(1)}(.))=\frac{1}{2} \max(\sum_i |\bra{i}
\varepsilon^{(1)}_{noise}(\rho) \ket{i}|) \leq 4C^2
\end{equation}
Note that to get the same amount of error for different systems, the
dimension of RF should increase proportional to the square of the
angular momentum of system.

After using the reference frame its state would be changed due to
back reaction effects.  Since the total $L_Z$ is conserved, if there
were no error in the implementation of a $Z$-inv unitary, the state
of the reference frame would be unchanged.  The probability of error
is less than $4C^2$, so with the probability of $1-4C^2$ or more the
reference frame stays in its initial state. To get a measure of how
the reference frame degrades, consider using it to implement a
unitary operation on a different system each time \cite{comment}.  After
using the reference frame $n$ times, as long as $nC^2 \ll 1$ the probability
of being in its initial state would be more than $1-4nC^2$ and so
the error would be less than $(1-4nC^2)4C^2$. We can see that the
state of the reference frame after the first use would be a mixture of
states from the set
$\{\ket{l_{RZ},m_{RZ}=l_{RZ}},\ket{l_{RZ},m_{RZ}=l_{RZ}-1},...,\ket{l_{RZ},m_{RZ}=l_{RZ}-2l_1}
\}$.

For  a  reference frame with fixed $l_{RZ}$,  $\ket{l_{RZ},l_{RZ}}$ seems to be
the best choice. However, taking into account lemma II and the
property of adding vectors that we have used in this scheme, we can
use other states of the form $\ket{l_{RZ},m_{RZ}=l_{RZ}-k}$ provided
$l_{RZ} \gg k^2$. It is straightforward to see in this case instead
of $4C^2$ the error would be almost $4C^2(2k+1)$.

\section{Implementing arbitrary unitaries using $Z$-inv unitaries} \label{All}

In this section we show how to implement all arbitrary unitary time
evolutions on a physical system, whether or not they commute with
$L_z$, by using a $Z$-inv unitary time evolution  acting over that
physical system combined with an auxiliary reference frame X-RF. We
assume the angular momentum of X-RF, $l_{RX}$,  is large compared to
the maximum angular momentum  of the physical system under
consideration, $l_{sys}$. The idea is that changes in the angular
momentum of the system caused by unitary transformations on the
system are compensated for by making changes in X-RF. We choose the
initial state of X-RF to have a large uncertainty in $L_Z$, such
that increasing or decreasing its $L_Z$ leaves it nearly unchanged.

We therefore take the initial state of X-RF to be an equal
superposition of the form
\begin{equation}
\ket{R_X,0}=\frac{1}{\sqrt{2N+1}}(\ket{m=-N}+\ldots+\ket{m=N})
\end{equation}
where $N=l_{RX}-2l_{sys}$, so that the sum is the subset of magnetic
quantum numbers within this range.  This will afford us freedom to
compensate for changes in $L_z$ of the system by modifying the state
$\ket{R_X,0}$, which we shall denote by $\ket{R_X}$. Note that the
expectation values of $L_Z$ and $L_Y$ are zero for this state,
whereas the expectation value of $L_X$ is nonzero, which means that
this state is pointing in the X direction. Unlike  $\ket{Z-RF}$,
which is an eigenvector of $L_Z$, $\ket{R_X}$ is not an eigenvector
of $L_X$ and has completely different character. We shall also
deploy the following notation
\begin{equation}
\ket{R_X,n}=\frac{1}{\sqrt{2N+1}}(\ket{m=-N+n}+\ldots+\ket{m=N+n})
\end{equation}
which we refer to as a shift of  $\ket{R_X}$ by $n$.

Initially X-RF is in the state  $\ket{R_X}$ which has a large amount
of  uncertainty in $L_Z$. To implement unitaries on the system that
do not commute with  $L_Z$, we compensate for the change in $L_Z$ by
shifting the state of X-RF such that the total $L_Z$ remains
constant. Under this change the final state of X-RF is
dependent on the state of the system, thereby entangling them.
However if $N$ is sufficiently large, which means the initial state
has a large uncertainty in $L_Z$, these shifts change $\ket{X-RF}$
by only a small amount. For very large $N$ the state of system
remains unentangled with X-RF.

Our goal is to implement any arbitrary unitary transformation $U$
where
\begin{equation}
U \ket{m,\lambda}=\sum_{m',\lambda'} U_{(m',\lambda'),(m,\lambda)}
\ket{m',\lambda'}
\end{equation}
and where $\ket{m,\lambda}$ is a state of the system. To implement
$U$ we construct the following $Z$-inv unitary
\begin{equation}
\mathscr{U}=\bigoplus_M \mathscr{U}^{(M)}
\end{equation}
in which each $\mathscr{U}^{(M)}$ acts in the following manner
\begin{eqnarray}
\mathscr{U}^{(M)} \ket{m,\lambda} \ket{M-m} &=& \sum_{m',\lambda'}
U_{(m',\lambda'),(m,\lambda)} \ket{m',\lambda'}\ket{M-m'}
 \quad \      \mathrm{if}\ \ \  |M|\leq N+l_{sys}\nonumber \\
 &=& \ket{m,\lambda} \ket{M-m}  \qquad \qquad \qquad \qquad \qquad  \mathrm{if}\ \  |M|> N+l_{sys}
\end{eqnarray}
where $\ket{m,\lambda}$ is in the Hilbert space of the main system and $\ket{M-m}$ is in the Hilbert space of X-RF.  Using the
unitarity of $U$ it is straightforward to check that $\mathscr{U}$
is also a unitary operator that commutes with $L_Z$.

Now the effect of $U$ on the initial state
$\ket{m,\lambda}\ket{R_X}$ is
\begin{eqnarray}
\mathscr{U} \ket{m,\lambda}\ket{R_X}&=&\frac{1}{\sqrt{2N+1}} \sum_{M= -N+m}^{N+m}\ \mathscr{U}^{(M)} \ket{m} \ket{M-m} \nonumber\\
&=&\frac{1}{\sqrt{2N+1}} \sum_{M= -N+m}^{N+m}\ \  \sum_{m',\lambda'} U_{(m',\lambda'),(m,\lambda)} \ket{m',\lambda'}\ket{M-m'}\nonumber\\
&=& \frac{1}{\sqrt{2N+1}} \sum_{m',\lambda'} U_{(m',\lambda'),(m,\lambda)} \ket{m',\lambda'} \sum_{M= -N+m}^{N+m} \ket{M-m'}\nonumber\\
&=& \sum_{m',\lambda'} U_{(m',\lambda'),(m,\lambda)}
\ket{m',\lambda'} \ket{R_X,m-m'}
\end{eqnarray}
For an arbitrary state of the system $\ket{\Theta}=\sum_{m,\lambda}
\theta_{m,\lambda} \ket{m,\lambda}$ we  have
\begin{equation} \label{shifted}
\mathscr{U} \ket{\Theta}\ket{R_X}= \sum_{m,\lambda}
\theta_{m,\lambda}(\sum_{m',\lambda'}  U_{(m',\lambda'),(m,\lambda)}
\ket{m',\lambda'} \ket{R_X,m-m'})
\end{equation}

%\fig{10cm}{4cm}{overlap}
%{\small The state $\ket{\Gamma} $ is defined as the  overlap of
%$\ket{R_X,2l_{sys}}$ and $\ket{R_X,-2l_{sys}}$.}
%{fig:overlap}

\begin{figure}%[h!]
\begin{center}
\includegraphics[width=.5 \columnwidth]{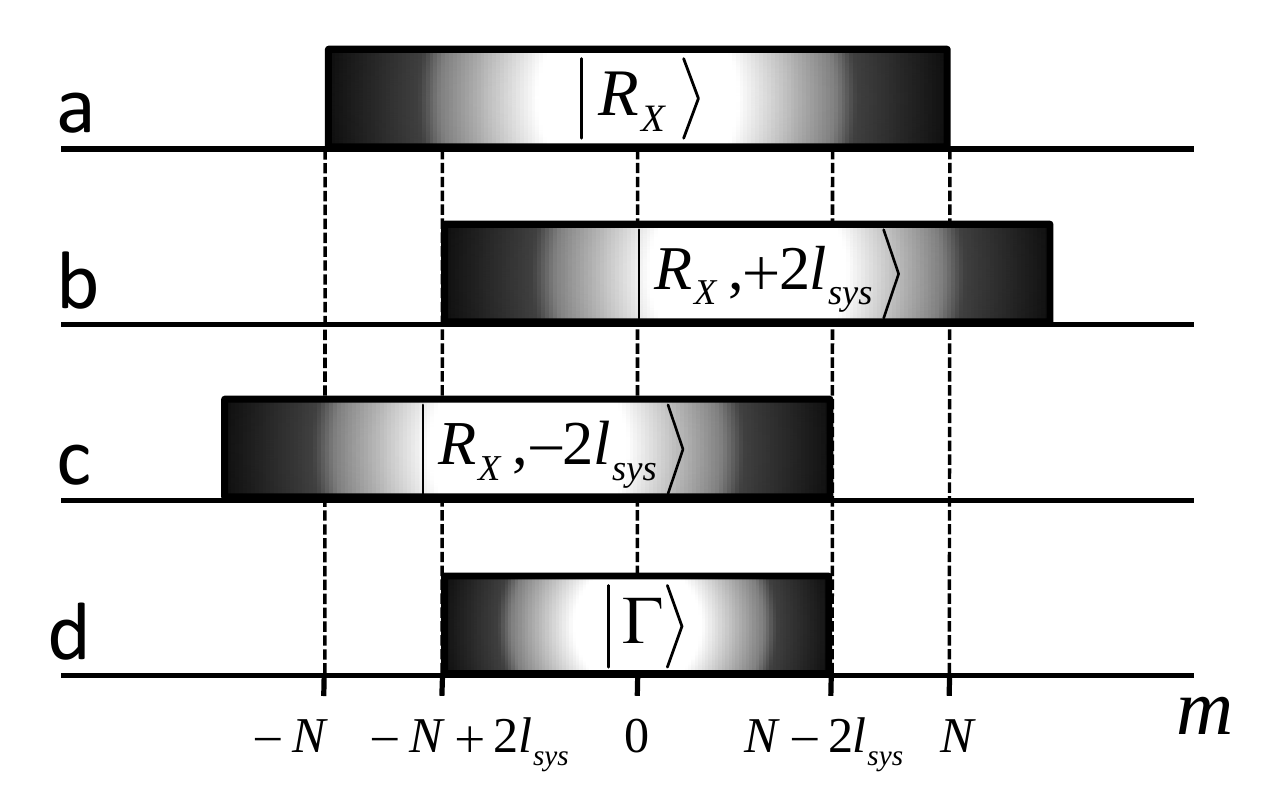}
\caption{ a) The state $\ket{R_X}$, b) the state $\ket{R_X,+2l_{sys}}$ which is the maximum shifted to the right of $\ket{R_X}$, c) the state $\ket{R_X,-2l_{sys}}$ which is the maximum shifted to the left of  $\ket{R_X}$ , d) the unnormalized vector  $\ket{\Gamma} $  which is the common part of all of these states. }
\end{center}
\label{over}
\end{figure}

%\begin{figure}
%\begin{center}
%\leavevmode
%\epsfxsize=4in
%\includegraphics[width=4.5in, height=3in]{overlap.eps}
%\epsfbox{overlap.eps}
%\end{center}
%\caption{ a) The state $\ket{R_X}$, b) the state $\ket{R_X,+2l_{sys}}$ which is the maximum shifted to the right of $\ket{R_X}$, c) the state $\ket{R_X,-2l_{sys}}$ which is the maximum shifted to the left of  %$\ket{R_X}$ , d) the unnormalized vector  $\ket{\Gamma} $  which is the common part of all of these states.}
%\label{add}
%\end{figure}

So  after time evolution the state of system becomes entangled with the state of the reference frame. This entangled state includes a superposition of vectors that are the tensor product of some state of the system and $\ket{R_X,m-m'}$, which is the initial state of the reference frame shifted by $m-m'$.  Since $-l_{sys}\leq
m \leq l_{sys}$ the maximum absolute value of these shifts is  $2l_{sys}$, yielding the extremal shifted states $\ket{R_X,-2l_{sys}}$ and $\ket{R_X,+2l_{sys}}$. We are interested in the largest common part between all of these shifted states. Therefore we  define the unnormalized vector $\ket{\Gamma}$  such that the overlap $\bra{\Gamma}{R_X,m-m'}\rangle$ is maximal and independent of
$\{m,m'\}$ for all allowed values of $\{m,m'\}$ (see fig. 2).  We can easily see that
\begin{equation}
\ket{\Gamma}=\sqrt{ \frac{1}{2N+1}  }
\sum_{i=-N+2l_{sys}}^{N-2l_{sys}}\ket{i}
\end{equation}
By writing
\begin{equation}
\ket{R_X,m-m'}=\ket{\Gamma}+ \left(\ket{R_X,m-m'}-
\ket{\Gamma}\right)
\end{equation}
we decompose $\ket{R_X,m-m'}$ as superposition of $\ket{\Gamma}$ and
another vector orthogonal to $\ket{\Gamma}$, a result that follows
from noting that
$$
\langle\Gamma\ket{R_X,m-m'} = {\langle\Gamma\ket{\Gamma}} =
\frac{2(N-2l_{sys})+1}{2N+1}
$$
Using this decomposition we obtain
\begin{equation}
\mathscr{U} \ket{\Theta}\ket{R_X}=U\ket{\Theta} \otimes
\ket{\Gamma} + \sum_{m,\lambda} \theta_{m,\lambda}\sum_{m',\lambda'}
U_{(m',\lambda'),(m,\lambda)} \ket{m',\lambda'} (\ket{R_X,m-m'}-
\ket{\Gamma})
\end{equation}
This describes the total state of the system and X-RF after this time
evolution. To find the effect of this time evolution on the system
we trace over the Hilbert space of X-RF. This yields a superoperator
that maps the initial state of system, $\rho$ to
\begin{equation}\label{e2}
\varepsilon^{(2)}(\rho)= \bra{\Gamma}\Gamma\rangle U\rho U^\dag +
(1-\bra{\Gamma}\Gamma\rangle) \varepsilon_{noise}^{(2)}(\rho)=
\frac{2(N-2l_{sys})+1}{2N+1} U\rho U^\dag + \frac{4l_{sys}}{2N+1}
\varepsilon_{noise}^{(2)}(\rho)
\end{equation}
where $\varepsilon_{noise}^{(2)}$ is a  trace-preserving completely
positive super-operator.  So using Eq.(\ref{distance2}) we see that
the error in the  outcome probability of any arbitrary measurement
is less than $d(U(.)U^\dag,\varepsilon^{(2)}) \leq
\frac{4l_{sys}}{2N+1}$. In the limit of large $N/l_{sys}$ the error
rate, $2l_{sys}/N$, is small.

\section{The First Scheme}

In section \ref{Z-inv-sec} we saw how  an $R$-inv time evolution
acting on the physical system and Z-RF can be used to implement any
arbitrary $Z$-inv unitary on the physical system. Then in section
\ref{All} we saw how a $Z$-inv unitary time evolution acting on the
main system and X-RF  can be used to implement any arbitrary unitary
on the physical system. By combining these two schemes we can
perform any arbitrary unitary on the system just by using $R$-inv
unitary time evolutions and Z-RF and X-RF (see fig. 3).

\begin{figure}%[h!]
\begin{center}
\includegraphics[width=.5 \columnwidth]{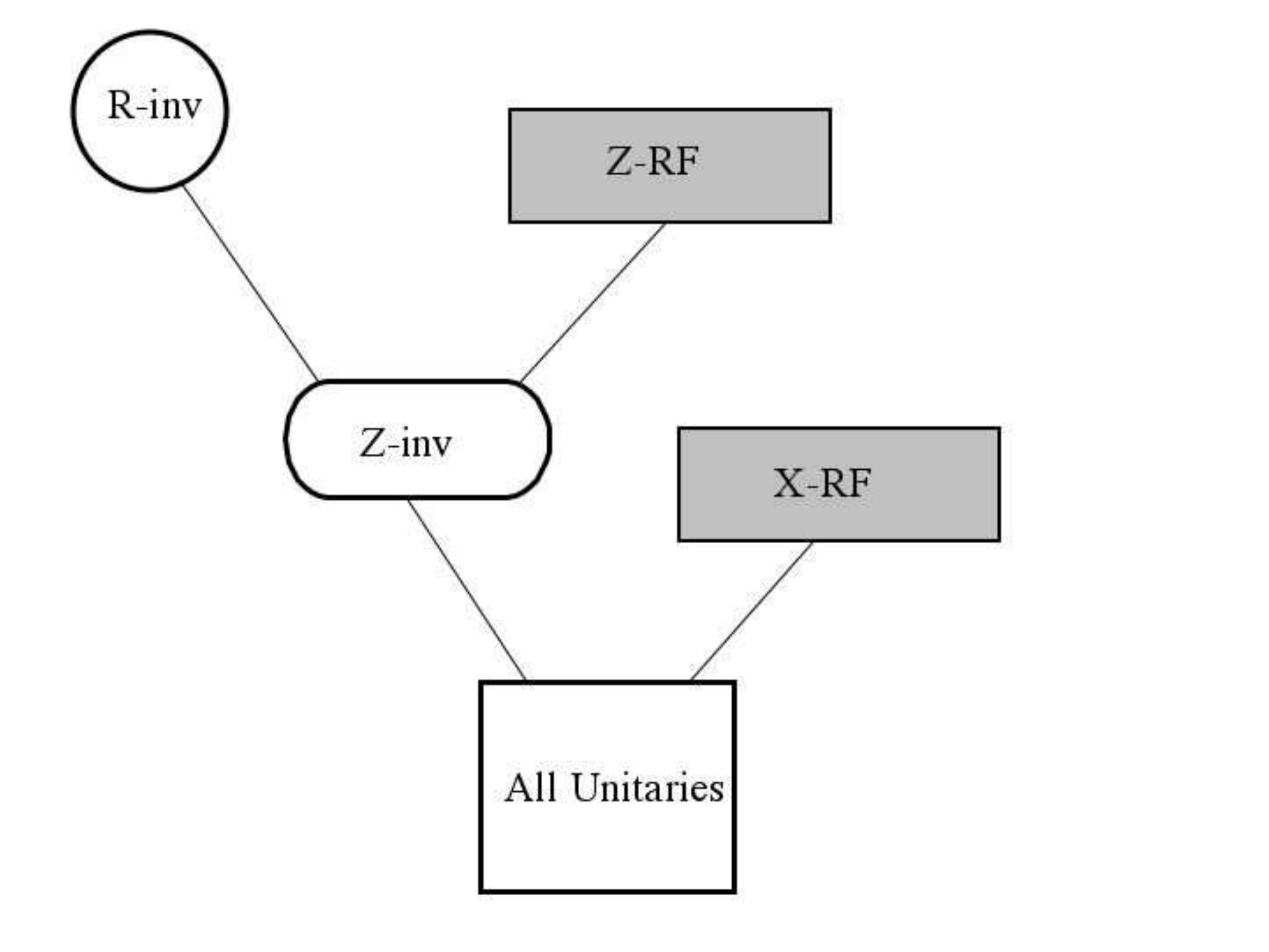}
\caption{ Schematic description of scheme 1.}
\end{center}
\label{schem1}
\end{figure}

%\fig{11cm}{8.5cm}{scheme1}
%{\small Schematic description of scheme 1.}
%{fig:schem1}

To implement an arbitrary unitary $U$ acting on the system, we need
to implement  the $Z$-inv  unitary $\mathscr{U}$ (defined in the
 previous section) on the system coupled to the auxiliary X-RF. To do so we let
the $R$-inv unitary  $\mathscr{V}=\sum_{j}   I_j \otimes \mathscr{U}^{(j-l_{RZ})}
$ act on the system coupled to both  X-RF and Z-RF.
Initially the density matrix is
\begin{equation}
\rho^{sys} \otimes \ket{R_X}\bra{R_X} \otimes \ket{Z-RF}\bra{Z-RF}
\end{equation}
After the time evolution the state of the system is
\begin{equation}
\varepsilon(\rho)=tr_{Z-RF,X-RF}(\mathscr{V}\rho^{sys} \otimes
\ket{R_X}\bra{R_X}  \otimes \ket{Z-RF}\bra{Z-RF}\mathscr{V}^\dag)
\end{equation}
$\mathscr{V}$ is designed to implement $\mathscr{U}$ on the physical
system and X-RF. So
\begin{equation}
\varepsilon(\rho)=tr_{X-RF}(\mathscr{U}\rho^{sys} \otimes
\ket{R_X}\bra{R_X}  \mathscr{U}^\dag)+
tr_{X-RF}(\varepsilon^{(1)}_{noise}(\rho^{sys} \otimes
\ket{R_X}\bra{R_X} ))
\end{equation}
From Eq.(\ref{e2}) we know that for large $N$
$$\varepsilon(\rho)=\frac{N-2l_{sys}}{N} U\rho U^\dag+ \frac{2l_{sys}}{N}
\varepsilon_{noise}^{(2)}(\rho)+tr_{X-RF}(\varepsilon^{(1)}_{noise}(\rho^{sys}
\otimes \ket{R_X}\bra{R_X}  ))$$
\begin{equation} =\frac{N-2l_{sys}}{N}
U\rho U^\dag + \varepsilon_{noise}(\rho) \label{rhonoise}
\end{equation}
where
\begin{equation}
\varepsilon_{noise}(\rho)= \frac{2l_{sys}}{N}
\varepsilon_{noise}^{(2)}(\rho)+tr_{X-RF}(\varepsilon^{(1)}_{noise}(\rho^{sys}
\otimes \ket{R_X}\bra{R_X}    ))
\end{equation}
To estimate the accuracy of this implementation we find an upper bound for  $d(U(.)U^\dag,\varepsilon(.))$
\begin{eqnarray}
d(U(.)U^\dag,\varepsilon(.))&=&\frac{1}{2} \max_{\rho,\{\ket{i}\}} \sum_i |\bra{i} U\rho U^\dag -\varepsilon(\rho)\ket{i}| \nonumber\\
&=&\frac{1}{2} \max_{\rho,\{\ket{i}\}}\sum_i  \left| \bra{i}\frac{2l_{sys}}{N} U\rho U^\dag -\frac{2l_{sys}}{N} \varepsilon_{noise}^{(2)}(\rho)\right. \nonumber\\
&&\qquad -\left. tr_{X-RF}(\varepsilon^{(1)}_{noise}(\rho^{sys}
\otimes   \ket{R_X}\bra{R_X} ))\ket{i}\right|
\end{eqnarray}

Using the triangle inequality and noting that
$\varepsilon_{noise}^{(2)}$ is positive and trace-preserving, we
find
\begin{eqnarray}
d(U(.)U^\dag,\varepsilon(.))&\leq& \frac{2l_{sys}}{N}+\frac{1}{2}
\max_{\rho,\{\ket{i}\}} \sum_i |\bra{i}
tr_{X-RF}(\varepsilon^{(1)}_{noise}(\rho^{sys} \otimes
\ket{R_X}\bra{R_X}   ))\ket{i}| \nonumber\\
&<& \frac{2l_{sys}}{N}+\frac{1}{2} \max_{\rho,\{\ket{i}\}} \sum_{i,s}
|\bra{i}\bra{s} \varepsilon^{(1)}_{noise}(\rho^{sys} \otimes
\ket{R_X}\bra{R_X}   )\ket{s}\ket{i}|
\end{eqnarray}
where $\{\ket{s}\}$ is orthogonal basis in the Hilbert space of
X-RF. Using Eq.(\ref{e1}) we can find an upper bound for the last
term in the right-hand side of the above equation. Since the angular
momentum of X-RF is $N+2l_{sys}$, the maximum total angular momentum
of the system and X-RF is $l_1=N+3l_{sys}$. However since
$N\gg l_{sys}$  we can assume $l_1 \approx N$. Thus using
Eq.(\ref{e1}) and noting that
$C^2=\frac{l_1^2+l_1+1/4}{2l_{RZ}}$ we obtain
\begin{equation}
d(U(.)U^\dag,\varepsilon(.))\leq \frac{2l_{sys}}{N}+
\frac{2N^2}{l_{RZ}}
\end{equation}
where $l_{RZ}\gg N$.
Obviously by choosing larger $l_{RZ}$ the error becomes
smaller. For any given $l_{RZ}$ there is a specific $N$ that minimizes this error.
After minimizing with respect to $N$ we have
\begin{equation} \label{error_first}
error_{I}(l_{sys},l_{RF})=
3\left(\frac{2l_{sys}^2}{l_{RZ}}\right)^{1/3}
\end{equation}
where $error_I$ shows the error in scheme I as a function of the
maximum angular momentum $l_{sys}$  of the system and the angular
momentum $l_{RF}$ of reference frame. We shall discuss the
implications of this result in section \ref{disc}.

\section{The Second Scheme}

In this section we propose an alternative method for building an
arbitrary unitary time evolution by using $R$-inv time evolutions.
As we have seen in section \ref{Z-inv-sec} we can implement all
$Z$-inv unitary evolutions by using $R$-inv unitary time evolutions
and a reference frame in the z direction. In the same way we can
build all unitaries commuting with $L_X$ ($X$-inv unitaries) by
using a reference frame X-RF that is defined in the same way as
Z-RF, but rotated so that the x direction is now the specified
direction. Note that the X-RF we use in this scheme differs from the
one we used in the first scheme. Now by a sequence of $Z$-inv and
$X$-inv unitaries we can build more unitary time evolutions. In
fact, as we will show in the following, from a sequence of unitaries
alternately commuting with $L_X$ and $L_Z$ we can build any
arbitrary unitary (see fig. 4).

\begin{figure}%[h!]
\begin{center}
\includegraphics[width=.85 \columnwidth]{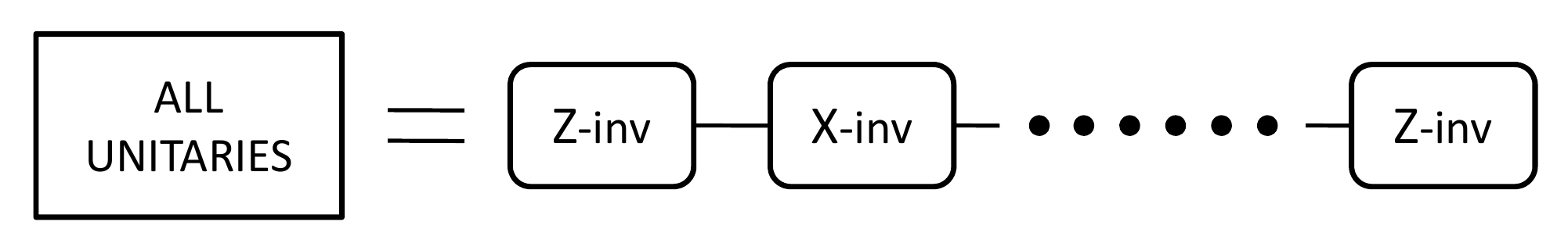}
\caption{ Schematic description of scheme 2: building all unitaries using a sequence of $Z-inv$ and $X-inv$ unitaries. }
\end{center}
\label{scheme2}
\end{figure}

%\begin{figure}
%\begin{center}
%\leavevmode
%\epsfxsize=4in
%\includegraphics[width=3.5in, height=0.6in]{scheme2.eps}
%\epsfbox{error.eps}
%\end{center}
%\caption{Error in scheme II as a function of $x$ where $x\propto l_{RF}/l^2_{sys}$ }
%\label{add}
%\end{figure}

%\fig{13cm}{2.5cm}{scheme2}
%{\small Schematic description of scheme 2.}
%{fig:scheme2}

The main idea is based on the following property using the four
Hamiltonians $\pm H_1$ and $\pm H_2$. By sequentially applying these
Hamiltonians we can construct any arbitrary Hamiltonian contained in
the Lie algebra generated by $\{H_1,H_2\}$. Consider the following
example. Apply $H_1$,  followed by $H_2$,$ -H_1$, and $-H_2$, each
for the same time
 $\delta t$. Since
\begin{equation}
e^{iH_1\delta t} e^{iH_2\delta t}e^{-iH_1\delta t}e^{-iH_2\delta
t}=e^{i(H_1H_2-H_2H_1)\delta t^2}+ O(\delta t^3)
\end{equation}
for small $\delta t$, the result is the same as if one had applied
$i[H_1,H_2]$ for time $\delta t ^2$. In general,  any given
Hamiltonian in the Lie algebra generated by $H_1$ and $H_2$, can
effectively be constructed by applying a sequence of  $\pm H_1$ and
$\pm H_2$ \cite{lloyd}.

Using $R$-inv interactions acting on the system and reference
frames, we are able to apply all Hamiltonians commuting with $L_Z$
and also all Hamiltonians commuting with $L_X$ by using  Z-RF and
X-RF respectively . The Lie algebra generated by these generators
describes the full set of Hamiltonians that can be constructed. The
following lemma illustrates how to find this Lie algebra.

\textbf{Lemma III} \emph{Suppose  $A$,$B$ are two Hermitian
operators with the property that no eigensubspace of $A$ is
orthogonal to any eigensubspace of $B$. Then the union of the set of
all unitaries commuting with $A$ and the set of all unitaries
commuting with $B$ is a universal set i.e. all unitary operators can
be constructed from a sequence of unitaries in those sets. Moreover
the length of required sequence  is uniformly bounded.}

This lemma is proven in the appendix. (It has a nice generalization
based on graph connectivity to an arbitrary number of operators
(instead of just $A$ and $B$) \cite{Graph}.) In any representation
$L_Z$ and $L_X$ have the property that no eigensubspace of $L_X$ is
orthogonal to an eigensubspace of $L_Z$. Consequently unitary
operators commuting with $L_Z$ and unitary operators commuting with
$L_X$  form a universal set.

Each time we use a reference frame it experiences some inevitable
backreaction \cite{degrade, Poulin_REF}.  For instance after performing a Z-inv time evolution on the system,  the
Z-RF is not in its initial state  $\ket{Z-RF}=\ket{l_{RZ},l_{RZ}}$ anymore, but instead is in a mixture of states including states with other magnetic quantum numbers.  This new state of Z-RF cannot specify the z direction as well as the initial state. So it cannot be corrected to get the initial state of Z-RF  by just using R-inv resources without using another Z-RF. We might employ this used Z-RF to perform the next Z-inv time evolution; however since this used Z-RF cannot specify z direction as well as the initial Z-RF we expect more noise in implementing the second Z-inv time evolution. Consequently to avoid increasing amounts of noise we need a
fresh reference frame after each use (or after several uses) when the Z-RF is degraded more than some specific threshold, dependent on the desired accuracy. Considering the reference frames as a resource it is important to know the number needed to implement a unitary time evolution with some precision. The length
of the required sequence is uniformly bounded, which means that
there exists an upper bound for the number of steps required to
build any arbitrary unitary.   How can we estimate this number?
In other words, how many times do we need to use those reference
frames?

To answer this question we employ the Solovay-Kitaev theorem.
According to this theorem if $G$ is a universal set of unitary
operators which produce a dense subset of $SU(d)$, and $G$ is closed
under inverse then $\forall U \in SU(d),\tau >0, \exists U_1,...,U_n
\in G $ such that $||U-U_1...U_n||\leq \tau $ and
$n=\mathscr{O}(\ln^2(\frac{1}{\tau}))$ \cite{Nielsen}.

In the present case $G$ consists of unitaries
commuting with $L_Z$ and unitaries commuting with $L_X$.
Suppose $U_{apx}$ is the approximation of $U$ that is
obtained by this method after $n$ steps $U_{apx}=U_1...U_n $ where $||U-U_{apx}||\leq \tau $. According to the Solovay-Kitaev theorem we can assume
$n\approx A  \ln^2(\frac{1}{\tau})$ where $A$ is a constant.
Using Eq.(\ref{unitary distance}) and assuming  $\tau$ is small we find $d(U(.)U^\dag,U_{apx}(.)U_{apx}^\dag)\leq \tau$.

Due to the finite size of the reference
frame we cannot perform $U_1,...,U_n$ perfectly;  each time there is an
error less than $4C^2$. So instead of
$U_1...U_n(.)U^\dag_n...U_1^\dag$ we implement
$\varepsilon(.)=\varepsilon_1(...\varepsilon_n(.))$. Using the
triangle inequality we have
\begin{equation}
d(U(.)U^\dag,\varepsilon(.))\leq
d(U(.)U^\dag,U_{apx}(.)U_{apx}^\dag)+d(U_{apx}(.)U_{apx}^\dag,\varepsilon(.))
\leq \tau +4nC^2=\tau +4A\ C^2
\ln^2(\frac{1}{\tau})
\end{equation}
where Eq. (\ref{trace_distance}) was repeatedly used to bound $d(U_{apx}(.)U_{apx}^\dag,\varepsilon(.))$.

The angular momentum of the reference frame, which is proportional to
$1/C^2$,  and the required number of reference frames
$n$ both can be regarded as limiting resources. In terms of these
resources total  error is bounded by
$$e^{-\sqrt{n/A}}+4nC^2$$
Obviously increasing the angular momentum of the reference frame decreases $C^2$ and with it the error.
As $C^2 \to 0$  this error approaches zero. However given a fixed $C^2$,  what
is the minimum total error we can achieve under this restriction?
By minimizing the total error $\tau +4A\ C^2  \ln^2(\frac{1}{\tau})$
with respect to $\tau$ we can find the minimum accessible error
under this constraint. Defining $\tau=e^{-\sigma}$ and
$x=(8AC^2)^{-1}$ the minimizing condition is
\begin{equation}
x= \sigma e^{\sigma}  \Longrightarrow \sigma = W(x)
\end{equation}
where $W(x)$ is the  Lambert-W function \cite{Lambert}.
%\fig{14.5cm}{7.5cm}{error}
%{\small Error in scheme II as a function of $x$ where $x\propto
%l_{RF}/l^2_{sys}$.}
%{fig:error}
\begin{figure}
\begin{center}
\leavevmode
\epsfxsize=4in
\includegraphics[width=6in, height=3in]{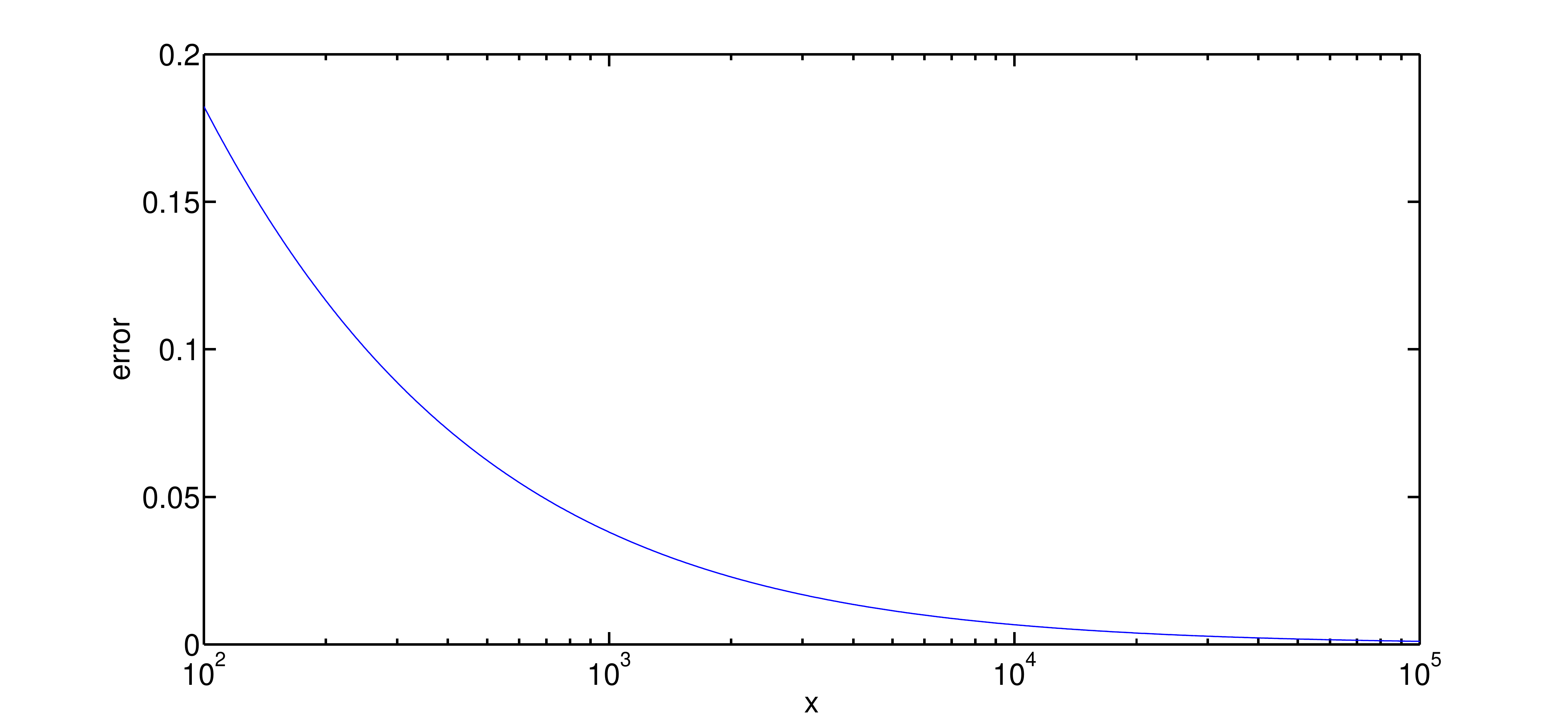}
%\epsfbox{error.eps}
\end{center}
\caption{Error in scheme II as a function of $x$ where $x\propto l_{RF}/l^2_{sys}$ }
\label{error}
\end{figure}
Hence the total error is bounded by
$$e^{-W(x)} +x^{-1} W^2(x)$$
(see fig.5) and the number of reference frames we need
would be $AW^2(x)$. Note that $x$ is equal to some constant times
$l_{RF}/l^2_{sys}$. For $x>1$ we can see that
$\ln(x)/2<W(x)<\ln(x)$. So at the limit of large $x$ the error is
bounded by
\begin{equation} \label{error_second}
error_{II}(l_{sys},l_{RF})=\ln^2(x)/x
\end{equation}
and the number of required reference frames grows like $\ln^2(x)$.

Finally we note that, though non-optimal, we can of course choose
any other direction instead of the x direction (except Z) for the
second reference frame. However a bad choice of independent
direction necessitates a longer sequence to reach an arbitrary
unitary.

\section{Discussion}\label{disc}

Associated with any restriction on resources is a
 theory describing whether under certain constraints
a given operation is feasible. Entanglement resource theory is a
well-known example. Analogously,  a
resource theory has more recently been developed for reference frames \cite{Resource}.
It is interesting to compare the two schemes proposed here from the viewpoint of the resources they require.

First note that the error in both of these schemes is a function of
$l_{sys}^2/l_{RF}$. Hence if the maximum angular momentum of the
system increases by a factor of $k$ then an increase in the angular
momentum of the reference frame by a factor of $k^2$  will yield the
same error. It is not clear whether there exist schemes in
which this factor is smaller than $k^2$.

Now let us check how the error decreases as the (large)
reference-frame angular momentum increases in these two schemes.
Looking to eqs. (\ref{error_first},\ref{error_second}) we see that
in the first scheme the reciprocal of the error grows as
$\mathscr{O}(\sqrt[3]{l_{RF}})$ while in the second scheme it grows
as $\mathscr{O}(l_{RF}/\ln^2(l_{RF}))$ . Recall that both these
schemes are based on a simpler scheme to implement $Z-inv$ unitary
time evolutions in which the reciprocal error is
$\mathscr{O}(l_{RF})$. So it seems reasonable that the reciprocal of
the error in both of these schemes grows slower than
$\mathscr{O}(l_{RF})$. A natural open problem is to determine the
best scheme for ensuring the error tends to zero as rapidly as
possible relative to $l_{RF}$.

 We
have shown that the presence of rotational symmetry does not
restrict us in performing arbitrary measurements or time evolutions
of the system. As there is evidently nothing specific in rotational
symmetry, we expect that the same holds for the existence of  other
symmetries. For example an extension of these results to more
complicated Lie groups, such as $SU(N)$, would be interesting. Here
the challenge is to construct an appropriate generalization of lemma
II.  Apart from this, there is nothing specific to SU(2) that would
appear to prohibit this kind of generalization.  Work on this issue
is in progress.

\vspace{1cm}

\textbf{Acknowledgments}

We are grateful to Lana Sheridan, Easwar Magesan and David Poulin for reading the manuscript and
providing important comments and introducing some
references. We thank Robert Spekkens for valuable discussions.
This work was supported by the Natural Sciences and Engineering
Research Council of Canada.

\vspace{1cm}
%\newpage
%
%\appendix

\section*{Appendix: Proofs of the Lemmas}

\subsection*{Lemma I:}
Suppose $\rho$ is a density operator and $O_1, O_2$ are two
arbitrary operators and $\{\ket{i}\}$ is an arbitrary set of an
orthogonal basis. Then
$$\sum_i |\bra{i}O_1 \rho O_2 \ket{i}| \leq ||O_1||\times ||O_2||$$

\noindent{\bf Proof}

Using the Cauchy-Schwartz inequality we have
\begin{equation}
\sum_i |\bra{i}O_1 \rho O_2 \ket{i}|=\sum_i |\bra{i}O_1 \sqrt{\rho}
\sqrt{\rho} O_2 \ket{i}|\leq \sqrt{\sum_i {\bra{i}O_1\rho O_1^\dag
\ket{i}}} \sqrt{\sum_i {\bra{i}O_2\rho O_2^\dag \ket{i}}}
\end{equation}

\begin{equation}
= \sqrt{tr(|O_1|^2\rho)tr(|O_2|^2\rho)}\leq ||O_1||\times ||O_2||
\end{equation}
Using this result we can get the more general result that
\begin{equation}
\sum_i |\bra{i}O_1\ldots O_k \rho O_{k+1}\ldots O_N \ket{i}| \leq
||O_1||\ldots ||O_N||
\end{equation}

\subsection*{Lemma II}
Consider the following state
\begin{equation}
\ket{\psi}=\ket{l_1,m_1}\ket{l_2,m_2=l_2-k}
\end{equation}
where $l_1,l_2$ are angular momentums and $m_1,m_2$ are eigenvalues
of $L_Z$. At the limit of  $l_2\gg l_1^2, k^2$ this state is almost
the same as the state with total angular momentum equal to $m_1+l_2$
and total $L_z$ equal to $m_1+l_2-k$ i.e.
\begin{equation}
\ket{\phi}=\ket{(j=m_1+l_2,m=m_1+l_2-k):(l_1,l_2)}
\end{equation}
or more precisely
\begin{equation}
|\bra{\phi}\psi\rangle|^2 \geq 1-\frac{(l_1^2+l_1-m_1^2)(2k+1)-m_1
}{2l_2}
\end{equation}
In terms of Clebsch-Gordon coefficients this means
$$lim_{l_1^2/l_2, k^2/l_2\rightarrow 0} \left( C_{l_2,l_2-k;l_1,m_1}^{m_1+l_2-k,m_1+l_2}\right)^2
= 1-\frac{(l_1^2+l_1-m_1^2)(2k+1)-m_1 }{2l_2} $$

\noindent{\bf Proof}

Define $m_{tot}=m_1+l_2-k$. For $i\geq m_{tot}$ we denote
\begin{equation}
\ket{i}\equiv \ket{(j=i,m=m_{tot}):(l_1,l_2)}
\end{equation}
Using this notation we can expand $\ket{\psi}$ as
\begin{equation}
\ket{\psi}=\alpha \ket{\phi}+ \sum_{i=m_{tot},i\neq
l_{tot}}^{l_1+l_2} \beta_i \ket{i}
\end{equation}
where $l_{tot}=m_1+l_2$ and $\ket{\phi}$, which is defined in the
lemma is actually the state $\ket{(j=l_{tot},m=m_{tot}):(l_1,l_2)}$.
Note that $L_z\ket{\psi}=m_{tot}\ket{\psi}$ and so vector states
with total angular momentum less than $m_{tot}$ do not appear in
this expansion. Using this expansion we have
\begin{equation}
J^2\ket{\psi}=\alpha l_{tot}(l_{tot}+1)\ket{\phi} +
\sum_{i=m_{tot},i\neq l_{tot}}^{l_1+l_2} i(i+1) \beta_i \ket{i}
\end{equation}
\begin{equation}
=\alpha l_{tot}(l_{tot}+1)\ket{\phi} + \sum_{i=m_{tot},i\neq
l_{tot}}^{l_1+l_2} l_{tot}(l_{tot}+1) \beta_i
\ket{i}+\sum_{i=m_{tot},i\neq l_{tot}}^{l_1+l_2}
[i(i+1)-l_{tot}(l_{tot}+1)] \beta_i \ket{i}
\end{equation}
\begin{equation} \label{J2}
=l_{tot}(l_{tot}+1)\ket{\psi}+ \sum_{i=m_{tot},i\neq
l_{tot}}^{l_1+l_2} [i(i+1)-l_{tot}(l_{tot}+1)] \beta_i \ket{i}
\end{equation}
On the other hand, using the relation
\begin{equation}
J^2=L_1^2+L_2^2+2\overrightarrow{L}_1.\overrightarrow{L}_2=L_1^2+L_2^2+2L_{1z}L_{2z}+L_{1+}L_{2-}+L_{1-}L_{2+}
\end{equation}
we obtain

$$J^2\ket{\psi}=[l_1(l_1+1)+l_2(l_2+1)+2m_1(l_2-k)]\ket{\psi}+$$
\begin{equation}
\sqrt{(l_1+m_1+1)(l_1-m_1)}\sqrt{(2l_2-k)(1+k)}\ket{\psi_1^\bot}+\sqrt{(l_1-m_1+1)(l_1+m_1)}\sqrt{k(2l_2-k+1)}\ket{\psi_2^\bot}
\end{equation}
Where
$$\ket{\psi_1^\bot}=\ket{l_1,m_1+1} \otimes \ket{l_2,m_2=l_2-k-1}\ ,\ \  \ket{\psi_2^\bot}=\ket{l_1,m_1-1} \otimes \ket{l_2,m_2=l_2-k+1}$$
are orthogonal to $\ket{\psi}$.\\
Equating this result with Eq.(\ref{J2}) we obtain
\begin{equation}
\sum_{i=m_{tot},i\neq l_{tot}}^{l_1+l_2} [i(i+1)-l_{tot}(l_{tot}+1)]
\beta_i \ket{i}=A \ket{\psi} + B \ket{\psi_1^\bot}+D
\ket{\psi_2^\bot}
\end{equation}
where
$$A=l_1(l_1+1)-m_1(m_1+1)-2m_1k\, \ \ \
B=\sqrt{(l_1+m_1+1)(l_1-m_1)(2l_2-k)(1+k)}\ $$
\begin{equation}
D=\sqrt{(l_1-m_1+1)(l_1+m_1)k(2l_2-k+1)}
\end{equation}
So we deduce that
\begin{equation}\label{condition}
\sum_{i=m_{tot},i\neq l_{tot}}^{l_1+l_2}
[i(i+1)-l_{tot}(l_{tot}+1)]^2 |\beta_i|^2=A^2+B^2+D^2
\end{equation}
{Now  we find the lower bound for the left-hand side of
Eq.(\ref{condition}) for a fixed value of $\sum
|\beta_i|^2=1-\alpha^2$. The minimum of this expression occurs  when
all $\beta_i$ are zero except that one for which the factor
$[i(i+1)-l_{tot}(l_{tot}+1)]^2$ is minimum. For $k>0$ this minimum
occurs when $i=l_{tot}-1$. For $k=0$, $i$ is larger than $l_{tot}$
and so  this minimum happens for $i=l_{tot}+1$. Putting these
$\beta_i$s in Eq.(\ref{condition}) and noting that
$|\alpha|^2=1-\sum |\beta_i|^2$ we obtain }\begin{equation}
A^2+B^2+D^2=\sum_{i=m_{tot},i\neq l_{tot}}^{l_1+l_2}
[i(i+1)-l_{tot}(l_{tot}+1)]^2 |\beta_i|^2 \geq
(1-|\alpha|^2)|2l_{tot}|^2
\end{equation}
or equivalently
\begin{equation}
|\alpha|^2 \geq   1-\frac{(A^2+B^2+D^2)}{4|l_{tot}|^2}
\end{equation}
This inequality is always true for all $l_1,l_2,k$. Now we assume
$l_2\gg l_1^2,k^2$ and so we obtain
\begin{equation}
|\alpha|^2 \geq 1-\frac{(l_1^2+l_1-m_1^2)(2k+1)-m_1 }{2l_2}
\end{equation}

\subsection*{Lemma III}
 Suppose  $A$,$B$ are two Hermitian operators with the property that no eigensubspace of $A$ is orthogonal to any eigensubspace of $B$. Then the union of the set of all unitaries commuting with $A$ and the set of all unitaries commuting with $B$  is a universal set i.e. all unitary operators can be constructed from
a sequence of unitaries in those sets. Moreover the length of
the required sequence  is uniformly bounded.

\noindent{\bf Proof}

We define $\mathscr{H}$ to be the linear space spanned by all
Hamiltonians that can be constructed by a sequence of applying
Hamiltonians commuting with $A$ and Hamiltonians commuting with $B$.
Suppose $\{\ket{\alpha_i,\sigma_i}\}$ are the eigenvectors of $A$
with eigenvalue $\alpha_i$ and $\{\ket{\beta_j,\zeta_j}\}$ are
eigenvectors of $B$ with eigenvalue $\beta_j$ where $\sigma_i$ and
$\zeta_j$ shows possible degeneracies. From the assumption of this lemma we
know that all of the following operators are in  $\mathscr{H}$.
 $$\{\ket{\alpha_i,\sigma_i}\bra{\alpha_i,\sigma'_i} \}\bigcup \{\ket{\beta_j,\zeta_j}\bra{\beta_j,\zeta'_j} \}$$
Moreover we know that $\mathscr{H}$ is closed under commutation. So
the following operators is also in $\mathscr{H}$
\begin{equation}
[\ket{\alpha_i,\sigma_i}\bra{\alpha_i,\sigma_i},\ket{\beta_j,\zeta_j}\bra{\beta_j,\zeta_j}]=
\bra{\alpha_i,\sigma_i}\beta_j,\zeta_j\rangle
\ket{\alpha_i,\sigma_i}\bra{\beta_j,\zeta_j}-\bra{\beta_j,\zeta_j}\alpha_i,\sigma_i\rangle
\ket{\beta_j,\zeta_j}\bra{\alpha_i,\sigma_i}
\end{equation}
From the assumptions of the lemma, we also know that for each pair of
$\alpha_i,\beta_j$ there exist some $\sigma_i,\zeta_j$  such that
$\bra{\alpha_i,\sigma_i}\beta_j,\zeta_j\rangle \neq 0$ and so the
following operator is a member of $\mathscr{H}$.
\begin{equation}
e^{i\theta} \ket{\alpha_i,\sigma_i}\bra{\beta_j,\zeta_j}-
e^{-i\theta} \ket{\beta_j,\zeta_j}\bra{\alpha_i,\sigma_i}
\end{equation}
where $e^{i\theta}\equiv
\bra{\alpha_i,\sigma_i}\beta_j,\zeta_j\rangle/
|\bra{\alpha_i,\sigma_i}\beta_j,\zeta_j\rangle|$. Also the
commutator of this operator with
$\ket{\beta_j,\zeta_j}\bra{\beta_j,\zeta'_j}+\ket{\beta_j,\zeta'_j}\bra{\beta_j,\zeta_j}$
is a member of $\mathscr{H}$

$$
[\ e^{i\theta} \ket{\alpha_i,\sigma_i}\bra{\beta_j,\zeta_j}-
e^{-i\theta} \ket{\beta_j,\zeta_j}\bra{\alpha_i,\sigma_i}\ ,\
\ket{\beta_j,\zeta_j}\bra{\beta_j,\zeta'_j}+\ket{\beta_j,\zeta'_j}\bra{\beta_j,\zeta_j}\
]$$
$$=e^{i\theta} \ket{\alpha_i,\sigma_i}\bra{\beta_j,\zeta'_j}+e^{-i\theta} \ket{\beta_j,\zeta'_j}
 \bra{\alpha_i,\sigma_i} +c_1\ket{\beta_j,\zeta_j}\bra{\beta_j,\zeta'_j}+c_1^*\ket{\beta_j,\zeta'_j}\bra{\beta_j,\zeta_j}+c_2\ket{\beta_j,\zeta_j} \bra{\beta_j,\zeta_j}$$
where $c_1$ is a complex number and $c_2$ is real and moreover we
have assumed $\zeta_j\neq \zeta'_j$. But the terms with coefficients
$c_1$ and $c_2$ are members of $\mathscr{H}$ and $\mathscr{H}$ is
closed under linear combination. So we deduce that $e^{i\theta}
\ket{\alpha_i,\sigma_i}\bra{\beta_j,\zeta'_j}+e^{-i\theta}
\ket{\beta_j,\zeta'_j} \bra{\alpha_i,\sigma_i} $ is also a member of
$\mathscr{H}$.  On the other hand, $\{\ket{\beta_j,\zeta_j}\}$ is a
complete basis and so we can expand all
$\{\ket{\alpha_i,\sigma_i}\}$ in terms of them. This implies that
that for all $\alpha_{i},\alpha_{j} $ and $\sigma_{i},\sigma_{j}$,
the operator $
\ket{\alpha_i,\sigma_i}\bra{\alpha_j,\sigma_j}+\ket{\alpha_j,\sigma_j}
\bra{\alpha_i,\sigma_i} $ is also a member of $\mathscr{H}$. So all
symmetric operators are in $\mathscr{H}$. Moreover we know that the
commutator of this operator and $\ket{\alpha_i,\sigma_i}
\bra{\alpha_i,\sigma_i}$ are also members of $\mathscr{H}$. This
would be
\begin{equation}
[\ket{\alpha_i,\sigma_i}\bra{\alpha_j,\sigma_j}+\ket{\alpha_j,\sigma_j}
\bra{\alpha_i,\sigma_i},\ket{\alpha_i,\sigma_i}
\bra{\alpha_i,\sigma_i}]= \ket{\alpha_j,\sigma_j}
\bra{\alpha_i,\sigma_i}-\ket{\alpha_i,\sigma_i}\bra{\alpha_j,\sigma_j}
\end{equation}
 So all asymmetric operators are also in $\mathscr{H}$ and therefore  $\mathscr{H}$ is equivalent to the space of all operators. This means that by a sequence of unitary time evolutions  commuting with  $A$ and unitary evolutions commuting with $B$ we can perform all unitaries.

It was shown in \cite{Uniform, uniform2} that if a compact Lie
algebra is generated by $\{X_1, X_2,\ldots, X_n \}$ then any member
of the associated Lie group can be generated by a sequence as $e^{X
t_1} e^{X t_2}... e^{X t_i}$  where for each of these exponentials
$X$ is a different member of  $\{X_1, X_2,\ldots, X_n \}$. Moreover
the length of this sequence is uniformly bounded. Since $U(N)$ is
compact we deduce that the length of the sequence we need to
generate  all members of $U(N)$ by a sequence of unitaries commuting
with $A$ and unitaries commuting with $B$ is uniformly bounded.


\begin{thebibliography}{99}
\bibitem{comment} Note that
if the RF is used repeatedly to control the same system errors may
add coherently. In the worst case this leads to an error rate that
is quadratically faster than incoherent noise.
\end{thebibliography}
\end{document}